\documentclass[conference]{IEEEtran}
\IEEEoverridecommandlockouts
\usepackage{cite}
\usepackage{amsmath,amssymb,amsfonts}
\usepackage{algorithmic}
\usepackage{graphicx}
\usepackage{textcomp}
\usepackage{xcolor}
\def\BibTeX{{\rm B\kern-.05em{\sc i\kern-.025em b}\kern-.08em
    T\kern-.1667em\lower.7ex\hbox{E}\kern-.125emX}}

\usepackage[square,numbers]{natbib}

\def\SCC{Specified Certainty Classification (SCC) 
  \gdef\SCC{SCC}}
\def\Thisfile{acronyms.tex}
\def\Thisfiledate{2021/08/13}
\def\NISS{National Institute of Statistical Sciences (NISS)
    \gdef\NISS{NISS}}

\def\NCSU{North Carolina State University (NCSU)
    \gdef\NCSU{NCSU}}
\def\UNC{University of North Carolina at Chapel Hill (UNC)
    \gdef\UNC{UNC}}
\def\CMU{Carnegie Mellon University (CMU)
    \gdef\CMU{CMU}}
\def\DU{Duke University (Duke)
    \gdef\DU{Duke}}
\def\UMI{University of Michigan (UMI)
    \gdef\UMI{UMi}}
\def\UMD{University of Maryland College Park (UMD)
    \gdef\UMD{UMd}}
\def\PU{Purdue University (Purdue)
    \gdef\PU{Purdue}}
\def\SMU{Southern Methodist University (SMU)
    \gdef\SMU{SMU}}
\def\GMU{George Mason University (GMU)
    \gdef\GMU{GMU}}
\def\UIC{University of Illinois at Chicago (UIC)
    \gdef\UIC{UIC}}
\def\LANL{Los Alamos National Laboratory (LANL)
    \gdef\LANL{LANL}}
\def\PNNL{Pacific Northwest National Laboratory (PNNL)
    \gdef\PNNL{PNNL}}
\def\GM{General Motors (GM)
    \gdef\GM{GM}}
\def\GSK{GlaxoSmithKline (GSK)
    \gdef\GSK{GSK}}
\def\VI{Visual Insights (VI)
    \gdef\VI{VI}}
\def\EIA{Energy Information Administration (EIA)
    \gdef\EIA{EIA}}
\def\EPA{Environmental Protection Agency (EPA)
    \gdef\EPA{EPA}}
\def\NCES{National Center for Education Statistics (NCES)
    \gdef\NCES{NCES}}
\def\BTS{Bureau of Transportation Statistics (BTS)
    \gdef\BTS{BTS}}
\def\BLS{Bureau of Labor Statistics (BLS)
    \gdef\BLS{BLS}}
\def\NCHS{National Center for Health Statistics (NCHS)
    \gdef\NCHS{NCHS}}
\def\BC{Census Bureau (Census)
    \gdef\BC{Census}}
\def\CB{Census Bureau (Census)
    \gdef\CB{Census}}
\def\NASS{National Agricultural Statistics Service (NASS)
    \gdef\NASS{NASS}}
\def\NSF{National Science Foundation (NSF)
    \gdef\NSF{NSF}}
\def\DMS{Division of Mathematical Sciences (DMS)
    \gdef\DMS{DMS}}
\def\CISE{Computer and Information Sciences and Engineering 
    (CISE)\gdef\CISE{CISE}}
\def\CATS{Committee on Theoretical and Applied Statistics (CATS)
    \gdef\CATS{CATS}}
\def\NRC{National Research Council (NRC)
    \gdef\NRC{NRC}}
\def\CNSTAT{Committee on National Statistics (CNSTAT)
    \gdef\CNSTAT{CNSTAT}}
\def\DOD{US Department of Defense (DoD)
    \gdef\DOD{DoD}}
\def\USGS{US Geological Survey (USGS)
    \gdef\USGS{USGS}}
\def\OMB{Office of Management and Budget (OMB)
    \gdef\OMB{OMB}}
\def\NSA{National Security Agency (NSA)
    \gdef\NSA{NSA}}
\def\DHS{Department of Homeland Security (DHS)
    \gdef\DHS{DHS}}
\def\CDC{Centers for Disease Control and Prevention (CDC)
    \gdef\CDC{CDC}}
\def\DARPA{Defense Advanced Research Projects Agency (DARPA)
    \gdef\DARPA{DARPA}}
\def\DOE{Department of Energy (DOE)
    \gdef\DOE{DOE}}
\def\CIA{Central Intelligence Agency (CIA)
    \gdef\CIA{CIA}}
\def\DTRA{Defense Threat Reduction Agency (DTRA)
    \gdef\DTRA{DTRA}}
\def\NIST{National Institute of Standards and Technology (NIST)
    \gdef\NIST{NIST}}
\def\NIAAA{National Institute on Alcohol Abuse and Alcoholism
    (NIAAA)
    \gdef\NIAAA{NIAAA}}
\def\ARO{Army Research Office (ARO)
    \gdef\ARO{ARO}}
\def\FDA{Food and Drug Administration (FDA)
    \gdef\FDA{FDA}}
\def\SAMSI{Statistical and Applied Mathematical Sciences
    Institute (SAMSI)\gdef\SAMSI{SAMSI}}

\def\NCDOT{North Carolina Department of Transporation (NCDOT)
    \gdef\NCDOT{NCDOT}}
\def\NCGBC{North Carolina Bioinformatics and Genomics Consortium 
    (NCGBC)\gdef\NCGBC{NCGBC}}
\def\RTI{RTI International (RTI)
    \gdef\RTI{RTI}}
\def\CIIT{CIIT Centers for Health Research (CIIT)
    \gdef\CIIT{CIIT}}
\def\DQRI{Data Quality Research Institute (DQRI)
    \gdef\DQRI{DQRI}}
\def\DIMACS{Center for Discrete Mathematics and Theoretical %
    Computer Science (DIMACS)\gdef\DIMACS{DIMACS}}
\def\HDF{Hereditary Disease Foundation (HDF)
    \gdef\HDF{HDF}}
\def\NCDM{National Center for Data Mining (NCDM)
    \gdef\NCDM{NCDM}}
\def\RTP{Research Triangle Park (RTP)
    \gdef\RTP{RTP}}
\def\ITDB{Intermodal Transportation Database (ITDB)
    \gdef\ITDB{ITDB}}
\def\TRI{Toxic Release Inventory (TRI)
    \gdef\TRI{TRI}}
\def\CPS{Current Population Survey (CPS)
    \gdef\CPS{CPS}}
\def\SASS{Schools and Staffing Survey (SASS)
    \gdef\SASS{SASS}}
\def\ITR{Information Technology Research (ITR)
    \gdef\ITR{ITR}}
\def\DC{data confidentiality (DC)
    \gdef\DC{DC}}
\def\DQ{data quality (DQ)
    \gdef\DQ{DQ}}
\def\DI{data integration (DI)
    \gdef\DI{DI}}
\def\IT{information technology (IT)
    \gdef\IT{IT}}
\def\SDL{statistical disclosure limitation (SDL)
    \gdef\SDL{SDL}}
\def\IQ{information quality (IQ)
    \gdef\IQ{IQ}}
\def\MCMC{Markov chain Monte Carlo (MCMC)
    \gdef\MCMC{MCMC}}
\def\CSV{comma-separated value (CSV)
    \gdef\CSV{CSV}}

\def\RMI{remote method invocation (RMI)
    \gdef\RMI{RMI}}
\def\SOAP{simple object access protocol (SOAP)
    \gdef\SOAP{SOAP}}
\def\XML{extensible markup language (XML)
    \gdef\XML{XML}}

\def\NHTSA{National Highway Traffic Safety Administration (NHTSA)%
    \gdef\NHTSA{NHTSA}}
\def\RDB{relational database (RDB)
    \gdef\RDB{RDB}\gdef\RDBMS{RBDMS}\gdef\RDBMSs{RDBMSs}}
\def\RDBMS{relational database management system (RDBMS)
    \gdef\RDB{RDB}\gdef\RDBMS{RBDMS}\gdef\RDBMSs{RDBMSs}}
\def\RDBMSs{relational database management systems (RDBMSs)
    \gdef\RDB{RDB}\gdef\RDBMS{RBDMS}\gdef\RDBMSs{RDBMSs}}
\def\GIS{geographical information system (GIS)\gdef\GIS{GIS}}
\def\DQTK{data quality toolkit (DQTK)
    \gdef\DQTK{DQTK}}
\def\DQRC{data quality report card (DQRC)
    \gdef\DQRC{DQRC}}
\def\TDQM{Total Data Quality Management (TQDM)
    \gdef\TDQM{TDQM}}
\def\OTR{optimal tabular release (OTR)
    \gdef\OTR{OTR}\gdef\OTRs{OTRs}}
\def\OTRs{optimal tabular releases (OTRs)
    \gdef\OTR{OTR}\gdef\OTRs{OTRs}}
\def\GUI{graphical user interface (GUI)
    \gdef\GUI{GUI}}
\def\OLTP{On-Line Transaction Processing (OLTP)
    \gdef\OLTP{OLTP}}
\def\GPRA{Government Performance Results Act (GPRA)
    \gdef\GPRA{GPRA}}

\def\CRM{Customer Relationship Management (CRM)
    \gdef\CRM{CRM}}
\def\HCI{human--computer interaction (HCI)
    \gdef\HCI{HCI}}
\def\NDHS{National Defense and Homeland Security (NDHS)
    \gdef\NDHS{NDHS}}
\def\MMR04{2004 Conference on Mathematical Methods in Reliability (MMR 2004)
    \gdef\MMR04{MMR 2004}}

\def\NCLB{No Child Left Behind Act (NCLB)
    \gdef\NCLB{NCLB}}
\def\CCD{Common Core of Data (CCD)
    \gdef\CCD{CCD}}

\def\NIH{National Institutes of Health (NIH)
    \gdef\NIH{NIH}}
\def\EFF{Electronic Frontier Foundation (EFF)
    \gdef\EFF{EFF}}
\def\GCD{graduation, completion and dropout (GCD)
    \gdef\GCD{GCD}}

\def\NCAR{National Center for Atmospheric Research (NCAR)
    \gdef\NCAR{NCAR}}

\def\TIMSS{Third International Mathematics and Science Study (TIMSS)
    \gdef\TIMSS{TIMSS}}
\def\PISA{Program for International Student Assessment (PISA)
    \gdef\PISA{PISA}}
\def\IEA{International Association for the Evaluation of
        Educational Achievement (IEA)
    \gdef\IEA{IEA}}
\def\NAEP{National Assessment of Educational Progress (NAEP)
    \gdef\NAEP{NAEP}}
\def\PIRLS{Progress in International Reading Literacy Study (PIRLS)
    \gdef\PIRLS{PIRLS}}

\def\OECD{Organization for Economic Cooperation and Development (OECD)
    \gdef\OECD{OECD}}

\def\GDC{graduation, dropout and completion (GDC)
    \gdef\GDC{GDC}}

\def\SMPC{secure multi-party computation (SMPC)
    \gdef\SMPC{SMPC}}
\def\HIPAA{Health Insurance Privacy and Accountability Act (HIPAA)
    \gdef\HIPAA{HIPAA}}

\def\ACS{American Community Survey (ACS)
    \gdef\ACS{ACS}}

\def\ESSI{Education Statistics Services Institute (ESSI)
    \gdef\ESSI{ESSI}}

\def\SCDM{Society for Clinical Data Management (SCDM)
    \gdef\SCDM{SCDM}}
\def\ACDM{Association for Clinical Data Management (ACDM)
    \gdef\ACDM{ACDM}}

\def\ASA{American Statistical Association (ASA)
    \gdef\ASA{ASA}}

\def\NCAA{National Collegiate Athletic Association (NCAA)
    \gdef\NCAA{NCAA}}
\def\GED{General Education Development (GED)
    \gdef\GED{GED}}

\def\ISU{Iowa State University (ISU)
    \gdef\ISU{ISU}}

\def\FIPS{Federal Information Processing System (FIPS)
    \gdef\FIPS{FIPS}}
\def\GSA{General Services Administration (GSA)
    \gdef\GSA{GSA}}

\def\AIR{American Institutes for Research (AIR)
    \gdef\AIR{AIR}}
\def\ESSIS{Education Statistics Services Institute---Statistics
    (ESSI--Stat)\gdef\ESSIS{ESSI-Stat}}

\def\NESSI{NAEP Education Statistics Services Institute (NESSI)
    \gdef\NESSI{NESSI}}

\def\ACM{Association for Computing Machinery (ACM)
    \gdef\ACM{ACM}}
\def\IEEE{Institute of Electrical and Electronics Engineers (IEEE)
    \gdef\IEEE{IEEE}}
\gdef\SIAM{Society for Industrial and Applied Mathematics (SIAM)
    \gdef\SIAM{SIAM}}
\def\IAOS{ISI Section on Official Statistics (IAOS)
    \gdef\IAOS{IAOS}}
\def\ISBA{International Society for Bayesian Analysis (ISBA)
    \gdef\ISBA{ISBA}}

\def\CDAC{Confidentiality and Data Access Committee (CDAC)
    \gdef\CDAC{CDAC}}
\def\CSIRO{Commonwealth Scientific and Industrial Research
    Organisation (CSIRO)\gdef\CSIRO{CSIRO}}
\def\TUCASI{Triangle Universities Center for Advanced Studies,
    Inc.\ (TUCASI)\gdef\TUCASI{TUCASI}}
\def\NCI{National Cancer Institute (NCI)
    \gdef\NCI{NCI}}

\def\BMSA{Board on Mathematical Sciences and their Applications
    (BMSA)\gdef\BMSA{BMSA}}

\def\NSCAW{National Survey of Child and Adolescent Well-Being
    (NSCAW)\def\NSCAW{NSCAW}}

\def\DAS{Data Analysis System (DAS)
    \gdef\DAS{DAS}\gdef\DASs{DAS's}}
\def\DASs{Data Analysis Systems (DAS's)
    \gdef\DAS{DAS}\gdef\DAS{DAS's}}

\def\CFFR{Committee on Federally Funded Research (CFFR)
    \gdef\CFFR{CFFR}}

\def\AJS{American Judicature Society (AJS)
    \gdef\AJS{AJS}}
\def\ECCR{Exploratory Center for Cheminformatics Research (ECCR)
    \gdef\ECCR{ECCR}}
\def\AAAS{American Association for the Advancement of Science (AAAS)
    \gdef\AAAS{AAAS}}

\def\RTF{Research Triangle Foundation (RTF)
    \gdef\RTF{RTF}}

\def\CFFR{Committee on Federally Funded Research (CFFR)%
    \gdef\CFFR{CFFR}}
\def\MPS{Directorate for Mathematical and Physical Sciences (MPS)%
    \gdef\MPS{MPS}}
\def\CMG{Collaborations in Mathematical Geosciences (CMG)%
    \gdef\CMG{CMG}}

\def\IMS{Institute of Mathematical Statistics (IMS)
    \gdef\IMS{IMS}}
\def\ISI{International Statistical Institute (ISI)
    \def\ISI{ISI}}
\def\IFNA{Interface Foundation of North America (IFNA)
    \gdef\IFNA{IFNA}}

\def\PPDM{privacy-preserving data mining (PPDM)
    \gdef\PPDM{PPDM}}

\def\ITSEW{International Total Survey Error Workshops (ITSEW)
    \gdef\ITSEW{ITSEW}}

\def\STEM{science, technology, engineering and mathematics (STEM)
    \gdef\STEM{STEM}}

\def\ASG{Art and Science Group (A\&SG)
    \gdef\ASG{A\&SG}}

\def\HSLS{High School Longitudinal Study (HSLS:09)
    \gdef\HSLS{HSLS:09}}
\def\NHIS{National Health Interview Survey (NHIS)
    \gdef\NHIS{NHIS}}
\def\FRG{Focused Research Groups in the Mathematical Sciences (FRG)
    \gdef\FRG{FRG}}

\def\IES{Institute of Education Sciences (IES)
    \gdef\IES{IES}}
\def\IOJ{Institute of Justice (IOJ)
    \gdef\IOJ{IOJ}}
\def\EAA{experimental analysis of algorithms (EAA)
    \gdef\EAA{EAA}\gdef\EAAC{EAA}}
\def\EAAC{Experimental analysis of algorithms (EAA)
    \gdef\EAA{EAA}\gdef\EAAC{EAA}}

\def\JPC{\textit{Journal of Privacy and Confidentialty} (JPC)
    \gdef\JPC{\textit{JPC}}}

\def\NWG{NISS Working Group (NWG)
    \gdef\NWG{NWG}\gdef\NWGS{NWGS}}
\def\NWGS{NISS Working Groups (NWGs)
    \gdef\NWGS{NWGS}\gdef\NWG{NWG}}

\def\DHHS{Department of Health and Human Services (DHHS)
    \gdef\DHHS{DHHS}}
\def\OCC{Office of the Comptroller of the Currency (OCC)
    \gdef\OCC{OCC}}

\def\FIPSE{Fund for the Improvement of Postsecondary Education
(FIPSE)
    \gdef\FIPSE{FIPSE}}

\def\FHWA{Federal Highway Administration (FHWA)\gdef\FHWA{FHWA}}

\def\SPAIG{Statistical Partnerships among Academia, Industry and
    Government (SPAIG)\gdef\SPAIG{SPAIG}}
\def\COPSS{Committee of Presidents of Statistical Societies (COPSS)
    \gdef\COPSS{COPSS}}

\def\SDDS{School District Demographics System (SDDS)
    \gdef\SDDS{SDDS}}

\def\ITRE{Institute for Transportation Research and Education (ITRE)
    \gdef\ITRE{ITRE}}
\def\TRB{Transportation Research Board (TRB)
    \gdef\TRB{TRB}}

\def\DTRA{Defense Threat Reduction Agency (DTRA)
    \def\DTRA{DTRA}}

\def\CPTAC{Clinical Proteomic Technology Assessment for Cancer (CPTAC)
    \gdef\CPTAC{CPTAC}}

\def\SRS{Division of Science Resources Statistics (SRS)
    \gdef\SRS{SRS}}
\def\SED{Survey of Earned Doctorates (SED)
    \gdef\SED{SED}}
\def\SDR{Survey of Doctorate Recipients (SDR)
    \gdef\SDR{SDR}}
\def\GSS{Survey of Graduate Students and Postdoctorates in
    Science and Engineering (GSS)
    \gdef\GSS{GSS}}
\def\RCG{National Survey of Recent College Graduates (RCG)
    \gdef\RCG{RCG}}
\def\NSCG{National Survey of College Graduates (NSCG)
    \gdef\NSCG{NCSG}}
\def\BRDIS{Business R\&D Innovation Survey (BRDIS)
    \gdef\BRDIS{BRDIS}}
\def\SANDE{science and engineering (S\&E)
    \gdef\SANDE{S\&E}}
\def\IPEDS{Integrated Postsecondary Education Data System (IPEDS)
    \gdef\IPEDS{IPEDS}}

\def\SEW{science, engineering and health workforce (SEHW)
    \gdef\SEW{SEHW}}

\def\CIPSEA{Confidential Information Protection and Statistical
    Efficiency Act of 2002 (CIPSEA)
    \gdef\CIPSEA{CIPSEA}}

\def\PSU{primary sampling unit (PSU)
    \gdef\PSU{PSU}\gdef\PSUS{PSUs}}
\def\PSUS{primary sampling units (PSUs)
    \gdef\PSU{PSU}\gdef\PSUS{PSUs}}

\def\NHANES{National Health and Nutrition Examination Survey (NHANES)
    \gdef\NHANES{NHANES}}
\def\ECLS{Early Childhood Longitudinal Study (ECLS)
    \gdef\ECLS{ECLS}}

\def\FERPA{Family Educational Rights and Privacy Act (FERPA)
    \gdef\FERPA{FERPA}}

\def\LBD{Longitudinal Business Database (LBD)
    \gdef\LBD{LBD}}
\def\LEHD{Longitudinal Employer-Household Dynamics (LEHD)
    \gdef\LEHD{LEHD}}
\def\SLDS{statewide longitudinal data systems (SLDS)
    \gdef\SLDS{SLDS}}

\def\ECLSK{Early Childhood Longitudinal Study--Kindergarten Class of
    1998--99 (ECLS-K)\gdef\ECLSK{ECLS-K}}
\def\SESTAT{Scientists and Engineers Statistical Data System (SESTAT)
    \gdef\SESTAT{SESTAT}}

\def\RENCI{Renaissance Computing Institute (RENCI)
    \gdef\RENCI{RENCI}}

\def\NIA{National Institute on Aging (NIA)
    \gdef\NIA{NIA}}
\def\SCOPE{Statistical Community of Practice and Engagement (SCOPE)
    \gdef\SCOPE{SCOPE}}
\def\ICSP{Interagency Council on Statistical Policy (ICSP)
    \gdef\ICSP{ICSP}}

\def\WSSM{World's Simplest Survey Microsimulator (WSSM)
    \gdef\WSSM{WSSM}}
\def\CES{Consumer Expenditure Survey (CES)
    \gdef\CES{CES}}

\def\NCSES{National Center for Science and Engineering Statistics (NCSES)
    \gdef\NCSES{NCSES}}

\def\TEP{Technical Expert Panel (TEP)
    \gdef\TEP{TEP}}
\def\ESSIN{Education Statistics Support Institute Network (ESSIN)
    \gdef\ESSIN{ESSIN}}
\def\FCSM{Federal Committee on Statistical Methodology (FCSM)
    \gdef\FCSM{FCSM}}
\def\JSM{Joint Statistical Meetings (JSM)
    \gdef\JSM{JSM}}
\def\NRBA{nonresponse bias analysis (NRBA)
    \gdef\NRBA{NRBA}}
\def\PSS{Private School Survey (PSS)
    \gdef\PSS{PSS}}
\def\CWI{comparable wage index (CWI)
    \gdef\CWI{CWI}}

\def\TSE{total survey error (TSE)
    \gdef\TSE{TSE}}

\def\TCRN{Triangle Census Research Network (TCRN)
    \gdef\TCRN{TCRN}}
\def\CER{comparative effectiveness research (CER)
    \gdef\CER{CER}}

\def\PII{personally identifiable information (PII)
    \gdef\PII{PII}}

\def\OES{Occupational Employment Statistics (OES)
    \gdef\OES{OES}}
\def\CPI{Consumer Price Index (CPI)
    \gdef\CPI{CPI}}
\def\CBSA{Core Based Statistical Area (CBSA)
    \gdef\CBSA{CBSA}\gdef\CBSAS{CBSAs}}
\def\CBSAS{Core Based Statistical Areas (CBSAs)
    \gdef\CBSA{CBSA}\gdef\CBSAS{CBSAs}}
\def\PUMA{Public Use Microdata Area (PUMA)
    \gdef\PUMA{PUMA}\gdef\PUMAS{PUMAs}}
\def\PUMAS{Public Use Microdata Areas (PUMAs)
    \gdef\PUMA{PUMA}\gdef\PUMAS{PUMAs}}
\def\PUMS{public use microdata samples (PUMS)
    \gdef\PUMS{PUMS}}

\def\RDC{Research Data Center (RDC)
    \gdef\RDC{RDC}\gdef\RDCS{RDCs}}
\def\RDCS{Research Data Centers (RDCs)
    \gdef\RDC{RDC}\gdef\RDCS{RDCs}}

\def\NCRN{NSF--Census Research Network (NCRN)
    \gdef\NCRN{NCRN}}
\def\NCRNCO{NSF--Census Research Network Coordination Office (NCRN-CO)
    \gdef\NCRNCO{NCRN-CO}}

\def\MPR{Mathematica Policy Research (MPR)
    \gdef\MPR{MPR}}
\def\NORC{NORC at the University of Chicago (NORC)
    \gdef\NORC{NORC}}
\def\AAPOR{American Association for Public Opinion Research (AAPOR)
    \gdef\AAPOR{AAPOR}}
\def\AEA{American Economic Association (AEA)
    \gdef\AEA{AEA}}

\def\SC{Steering Committee (SC)
    \gdef\SC{SC}}
\def\AEA{American Economic Association (AEA)
    \gdef\AEA{AEA}}
\def\ICES{International Conference on Establishment Statistics (ICES)
    \def\ICES{ICES}}
\def\IRS{Internal Revenue Service (IRS)
    \gdef\IRS{IRS}}
\def\SSA{Social Security Administration (SSA)
    \gdef\SSA{SSA}}
\def\COSSA{Consortium of Social Science Associations (COSSA)
    \gdef\COSSA{COSSA}}
\def\COPAFS{Council of Professional Associations on Federal Statistics (COPAFS)
    \gdef\COPAFS{COPAFS}}
\def\BJS{Bureau of Justice Statistics (BJS)
    \gdef\BJS{BJS}}
\def\ERS{Economic Research Service (ERS)
    \gdef\ERS{ERS}}
\def\USDA{U.S. Department of Agriculture (USDA)
    \gdef\USDA{USDA}}
\def\BEA{Bureau of Economic Analysis (BEA)
    \gdef\BEA{BEA}}

\def\NSO{national statistical office (NSO)
    \gdef\NSO{NSO}\gdef\NSOS{NSOs}}
\def\NSOS{national statistical offices (NSOs)
    \gdef\NSO{NSO}\gdef\NSOS{NSOs}}

\def\PI{Principal Investigator (PI)
    \def\PI{PI}}
\def\PIs{Principal Investigators (PIs)
    \def\PIs{PIs}}
\def\LEHD{Longitudinal Employer-Household Dynamics (LEHD)
    \def\LEHD{LEHD}}
\def\INSEE{Institut national de la statistique et des \'etudes \'economiques (INSEE)
    \def\INSEE{INSEE}}
\def\CREST{Centre de Recherche en \'Economie et Statistique (CREST)
    \def\CREST{CREST}}
\def\IAB{Institut f\"ur Arbeitsmarkt- und Berufsforschung (IAB)
     \def\IAB{IAB}}
\def\CBS{Centraal Bureau voor de Statistiek (CBS)
      \def\CBS{CBS}}
\def\MCU{multipoint conferencing units (MCU)
      \def\MCU{MCU}}

\def\CESCB{Center for Economic Studies (CES)
    \gdef\CESCB{CES}}

\def\SACNAS{Society for Advancement of Chicanos and Native Americans in Science (SACNAS)
    \gdef\SACNAS{SACNAS}}
\def\AWM{Association for Women in Mathematics (AWM)
    \gdef\AWM{AWM}}

\def\USPS{United States Postal Service (USPS)
    \gdef\USPS{USPS}}

\def\HSB{High School and Beyond (HS\&B:82)
    \gdef\HSB{HS\&B:82}}
\def\NELS{National Educational Longitudinal Study (NELS:88)
    \gdef\NELS{NELS:88}}
\def\NLS{National Longitudinal Study (NLS:72)
    \gdef\NLS{NLS:72}}
\def\NDI{National Death Index (NDI)
    \gdef\NDI{NDI}}

\def\NAS{National Academy of Sciences (NAS)
    \gdef\NAS{NAS}}
\def\EPRI{Electric Power Research Institute (EPRI)
    \gdef\EPRI{EPRI}}
\def\PCORI{Patient-Centered Outcomes Research Institute (PCORI)
    \gdef\PCORI{PCORI}}
\def\SHRP2{Strategic Highway Research Program 2 (SHRP 2)
    \def\SHRP2{SHRP 2}}
\def\OMOP{Observational Medical Outcomes Partnership (OMOP)
    \gdef\OMOP{OMOP}}
\def\ASM{Annual Survey of Manufactures (ASM)
    \gdef\ASM{ASM}}
\def\SIPP{Survey of Income and Program Participation (SIPP)
    \gdef\SIPP{SIPP}}

\def\NSRCG{National Survey of Recent College Graduates (NSRCG)
	\gdef\NSRCG{NSCRG}}
\def\SEH{science, engineering and health (SEH)
    \gdef\SEH{SEH}}
\def\HRS{University of Michigan Health and Retirement Study (HRS)
    \gdef\HRS{HRS}}
\def\NPSAS{National Postsecondary Student Aid Study (NPSAS)
    \gdef\NPSAS{NPSAS}}
\def\BnB{Baccalaureate and Beyond (B\&B)
    \gdef\BnB{B\&B}}
\def\BPS{Beginning Postsecondary Students Longitudinal Study (BPS)
    \gdef\BPS{BPS}}

\def\CMSS{Computation Methods in the Social Sciences (CMSS)
    \gdef\CMSS{CMSS}}

\def\TCS{Teacher Compensation Survey (TCS)
	\gdef\TCS{TCS}}
\def\SASS{Schools and Staffing Survey (SASS)
	\gdef\SASS{SASS}}
\def\NEA{National Education Association (NEA)
	\gdef\NEA{NEA}}
\def\FARS{Fatality Analysis and Reporting System (FARS)
	\gdef\FARS{FARS}}
\def\EDA{exploratory data analysis (EDA)
	\gdef\EDA{EDA}}

\def\CAT{computerized adaptive testing (CAT)
    \gdef\CAT{CAT}}
\def\CBT{computer-based testing (CBT)
    \gdef\CBT{CBT}}

\def\IRT{item response theory (IRT)
    \gdef\IRT{IRT}}

\def\CAPI{computer-assisted personal interview (CAPI)
    \gdef\CAPI{CAPI}}
\def\CATI{computer-assisted telephone interview (CATI)
    \gdef\CATI{CATI}}

\def\CAR{conditional autoregressive (CAR)
    \gdef\CAR{CAR}}

\def\IPUMS{Integrated Public Use Microdata Series (IPUMS)
    \gdef\IPUMS{IPUMS}}

\def\LEA{local education authority (LEA)
    \gdef\LEA{LEA}\gdef\LEAS{LEAs}}
\def\SEA{state education authority (SEA)
    \gdef\SEA{SEA}\gdef\SEAS{SEAs}}
\def\LEAS{local education authorities (LEAs)
    \gdef\LEA{LEA}\gdef\LEAS{LEAs}}
\def\SEAS{state education authorities (SEAs)
    \gdef\SEA{SEA}\gdef\SEAS{SEAs}}

\def\CMS{Centers for Medicare and Medicaid Services (CMS)
    \gdef\CMS{CMS}}

\def\MMS{Methodology, Measurement, and Statistics (MMS)
    \gdef\MMS{MMS}}

\def\SOC{standard occupational code (SOC)
    \gdef\SOC{SOC}}
\def\ASCO{American Society for Clinical Oncology (ASCO)
    \gdef\ASCO{ASCO}}

\def\UI{unemployment insurance (UI)
    \gdef\UI{UI}}

\def\PT{Project TALENT (PT)
    \gdef\PT{PT}}

\def\AERA{American Educational Research Association (AERA)
    \gdef\AERA{AERA}}

\def\CBSA{Core Business Statistical Area (CBSA)
    \gdef\CBSA{CBSA}\def\CBSAs{CBSAs}}
\def\CBSAs{Core Business Statistical Areas (CBSAs)
    \gdef\CBSA{CBSA}\def\CBSAs{CBSAs}}

\def\PT{Project Talent (PT)
    \gdef\PT{PT}}
\def\FCB{First Citizens Bank (FCB)
    \gdef\FCB{FCB}}
\def\BWF{Burroughs Wellcome Fund (BWF)
    \gdef\BWF{BWF}}

\def\PPRL{privacy-preserving record linkage (PPRL)
    \gdef\PPRL{PPRL}}

\def\NIA{National Institute on Aging (NIA)
    \gdef\NIA{NIA}}

\def\CHAID{Chi-squared automatic interaction detection (CHAID)
    \gdef\CHAID{CHAID}}
\def\GSC{General School Characteristics (GSC)
    \gdef\GSC{GSC}}

\def\CoDA{Center of Excellence for Complex Data Analysis (CoDA)
    \gdef\CoDA{CoDA}}

\def\AIC{Akaike information criterion (AIC)
    \gdef\AIC{AIC}}
\def\BIC{Bayes information criterion (BIC)
    \gdef\BIC{BIC}}
\def\SSES{Social, Statistical, and Environmental Sciences (SSES)
    \gdef\SSES{SSES}}
\def\SCSS{Survey, Computing, and Statistical Sciences (SCSS)
    \gdef\SCSS{SCSS}}
\def\DSDS{Division for Statistical and Data Science (DSDS)
    \gdef\DSDS{DSDS}}

\def\CM{Census of Manufactures (CM)
    \gdef\CM{CM}}

\def\FSRDC{Federal Statistical Research Data Center (FSRDC)
    \gdef\FSRDC{FSRDC}\gdef\FSRDCS{FSRDCs}}
\def\FSRDCS{Federal Statistical Research Data Centers (FSRDCs)
    \gdef\FSRDC{FSRDC}\gdef\FSRDCS{FSRDCs}}

\def\PTTP{partially trusted third party (PTTP)
    \gdef\PTTP{PTTP}\gdef\PTTPS{PTTPs}}
\def\PTTPS{partially trusted third parties (PTTPs)
    \gdef\PTTP{PTTP}\gdef\PTTPS{PTTPs}}

\def\NAICS{North American Industrial Classification System (NAICS)
    \gdef\NAICS{NAICS}}
\def\CART{classification and regression trees (CART)
    \gdef\CART{CART}}

\def\FEBRL{Freely Extensible Biomedical Record Linkage (FEBRL)
    \gdef\FEBRL{FEBRL}}
\def\FRIL{Fine-Grained Records Integration and Linkage Tool (FRIL)
    \gdef\FRIL{FRIL}}
\def\SOEMPI{Secure Open Enterprise Master Patient Index (SOEMPI)
    \gdef\SOEMPI{SOEMPI}}

\def\SDA{statistical disclosure avoidance (SDA)
    \gdef\SDA{SDA}}

\def\LAS{Laboratory for Analytic Sciences (LAS)
    \gdef\LAS{LAS}}

\def\IPF{iterative proportional fitting (IPF)
    \gdef\IPF{IPF}}
\def\ROC{receiver operating characteristic (ROC)
    \gdef\ROC{ROC}}

\def\NCBI{National Center for Biotechnology Information (NCBI)
    \gdef\NCBI{NCBI}}

\def\SCC{Specified Certainty Classification (SSC)
    \gdef\SCC{SCC}}

\def\BP{base pairs (BP) 
    \gdef\BP{BP}}

\def\ROC{receiver operating characteristic (ROC)
    \gdef\ROC{ROC}}

\gdef\MS{\texttt{Mason\_simulator}}

\gdef\MV{\texttt{Mason\_variator}}

\def\AI{artificial intelligence (AI)
   \gdef\AI{AI}}

\def\SARS{severe acute respiratory syndrome (SARS)
  \gdef\SARS{SARS}}

\def\SNP{single nucleotide polymorphism (SNP)
  \gdef\SNP{SNP}
  \gdef\SNPs{SNPs}}
  
\def\SNPs{single nucleotide polymorphisms (SNPs)
  \gdef\SNPs{SNPs}
  \gdef\SNP{SNP}}

\def\CRISPR{clustered regularly interspaced short palindromic repeats (CRISPR)
  \gdef\CRISPR{CRISPR}}

\def\CAS{CRISPR associated sequence (CAS)
  \gdef\CAS{CAS}}

\begin{document}

\title{Specified Certainty Classification,\\
with Application to Read Classification\\
for Reference-Guided Metagenomic Assembly
\thanks{This research was supported in part by NIH grant 5R01AI100947--06, ``Algorithms and Software for the Assembly of Metagenomic Data,'' to the University of Maryland College Park(Mihai Pop, PI).}}

\author{
\IEEEauthorblockN{Alan F. Karr}
\IEEEauthorblockA{\textit{Center Mid-Atlantic}}
\textit{Fraunhofer USA}\\
Riverdale, MD \\
akarr@fraunhofer.org
\and
\IEEEauthorblockN{Jason Hauzel}
\IEEEauthorblockA{\textit{Center Mid-Atlantic}}
\textit{Fraunhofer USA}\\
Riverdale, MD \\
jhauzel@fraunhofer.org
\and
\IEEEauthorblockN{Prahlad Menon}
\IEEEauthorblockA{\textit{Center Mid-Atlantic}}
\textit{Fraunhofer USA}\\
Riverdale, MD \\
pmenon@fraunhofer.org
\and
\IEEEauthorblockN{Adam A. Porter}
\IEEEauthorblockA{\textit{Department of}}
\textit{Computer Science}\\
\textit{University of Maryland}\\
College Park, MD \\
aporter@umd.edu
\and
\IEEEauthorblockN{Marcel Schaefer}
\IEEEauthorblockA{\textit{Center Mid-Atlantic}}
\textit{Fraunhofer USA}\\
Riverdale, MD \\
mschaefer@fraunhofer.org
}

\maketitle

\begin{abstract}
Specified Certainty Classification (SCC) is a new paradigm for employing classifiers whose outputs carry uncertainties, typically in the form of Bayesian posterior probabilities. By allowing the classifier output to be less precise than one of a set of atomic decisions, SCC allows all decisions to achieve a specified level of certainty, as well as provides insights into classifier behavior by examining all decisions that are possible. Our primary illustration is read classification for reference-guided genome assembly, but we demonstrate the breadth of SCC by also analyzing COVID-19 vaccination data.
\end{abstract}


\section{Introduction}\label{sec.introduction}
Classifiers are ubiquitous in bioinformatics, with applications ranging from genomics---the primary setting here---to radiology. A classifier,
developed from labeled training data, places new objects into one of a finite number of classes. Nearly all extant classifiers, including those based on deep learning models in \AI, provide little or no information regarding associated uncertainties \cite{friedmanhastietibshirani-2001}. Even more rarely is there control over those uncertainties. \emph{\SCC}, introduced here, gives analysts control over the certainty levels of the output of classifiers based on Bayesian posterior probabilities.

The distinguishing characteristics of \SCC\ are that: (1) All data point-specific decisions attain a specified level of certainty, denoted by $\beta$ below; (2) \SCC\ allows the analyst to investigate \emph{all decisions that can be made} rather than \emph{what decisions are made} for specific settings of the threshold; and (3) \SCC\ allows investigation of the thresholds at which data points lose precision. We discuss each briefly here and in detail below.

To achieve specified certainty, \SCC\ permits some decisions to be less precise than others. Specifically, if $\mathcal{A}$ is the set of atomic decisions, then rather than restrict output to be \emph{elements} of $\mathcal{A}$, \SCC\ allows a decision to be any nonempty \emph{subset} of $\mathcal{A}$. To illustrate, in \S\ref{sec.scc}, where the problem is to classify (short) DNA sequence reads as arising from one of three ``reference'' genomes---in this case, an adenovirus genome and two coronavirus genomes, $\mathcal{A} = \{\mathrm{Adeno}, \mathrm{COVID}, \mathrm{SARS}\}$, but some reads may be classified as ``COVID or SARS,'' or even ``Adeno or COVID or SARS.'' Loss of precision occurs when the decision is compound rather than atomic.\footnote{Thus use of ``precision'' is not that in machine learning, where precision means the same as positive predictive value in biostatistics.}

By investigating all decisions, \SCC\ embodies the paradigm in \cite{recordlinkage-plos1-2019}, where inpatient and outpatient databases from a major healthcare system in the U.S. were used to compare the performance of a number of software packages for record linkage. That paper demonstrated that some packages have dramatically richer decision-making capability than others. In turn, this capability facilitates construction of \ROC\ curves \cite{fawcett-roc-2016} that quantify the tradeoffs between false negatives (records that should have been linked, but were not, which is conventionally interpreted as bias) and false positives (records that were linked but should not have been, which is interpreted as noise), and enables construction of ensemble methods that outperform all the individual methods. Here, we are able to construct \ROC-like curves for \SCC\ that quantify the increase in accuracy resulting from loss in precision. See \S\ref{sec.scc}.

Transition points are particularly insightful for small datasets: in \S\ref{sec.alternatives} we apply \SCC\ in a healthcare context, to investigate COVID-19 vaccination rates in U.S. states.

The major contributions of the paper are formulation of \SCC\ and demonstration of its power in two very different bioinformatics/biomedicine contexts. The paper is organized as follows. We present \SCC\ concretely, via the read classification problem. \S\ref{sec.experiment} describes the experiment we conducted, and \S\ref{sec.claasifier} introduces the Bayesian classifier, which is based on triplet distributions \cite{markovstructure-2021} for the reference genomes. \S\ref{sec.scc} presents \SCC\ and articulates its properties. To demonstrate broad applicability, \S\ref{sec.alternatives} applies \SCC\ to COVID-19 vaccination data. Conclusions appear in \S\ref{sec.conclusion}.

\section{Experimental Context}\label{sec.experiment}
Our three reference genomes are an adenovirus (sometimes Adeno) genome of length 34,125, downloaded with the read simulator \texttt{Art}, a SARS-CoV-2 genome of length 29,926 contained in a coronavirus dataset downloaded from \NCBI\ in November of 2020,\footnote{The link is https://www.ncbi.nlm.nih.gov/datasets/coronavirus/genomes/.} which we call COVID, and a \SARS\ genome of length 29,751 from the same \NCBI\ database.

The \MS\ read simulator \cite{fu_mi_publications962} was used to simulate Illumina\footnote{Illumina is a major manufacturer of instruments for genome sequencing; their technology is optical in nature; see https://www.illumina.com/.} reads of length 101 from each of the three genomes, with approximately 6X coverage. The numbers of reads in our dataset are 1966, 1996 and 1907, respectively, which total to 5869. Parameters of \MS\ were set at their default values. The \MS\ introduces errors in the form of transpositions (SNPs), indels and undetermined bases, which following convention appear in the simulated reads as ``N'' and must be accommodated in computation of likelihood functions.

The read classification problem is to determine the source genome for each of the 5869 reads. All analyses reported below were performed using R \cite{R-2021}.

\section{The Bayesian Classifier}\label{sec.claasifier}
As in any Bayesian analyses, there are three components. The first is the reads themselves---the data. The second is the prior probabilities---for each read $R$, a probability distribution $\pi_R$ over $\mathcal{A} = \{\mathrm{Adeno}, \mathrm{COVID}, \mathrm{SARS}\}$, the set of atomic decisions. The third component comprises a likelihood function for each genome, calculated from the triplet distributions of nucleotides (bases) using the methodology outlined below and investigated at length in \cite{markovstructure-2021}. We refer to those as models, for which the three genomes comprise the training data. The likelihood functions are denoted by $L_A(\cdot)$, $L_C(\cdot)$ and $L_S(\cdot)$ for adenovirus, COVID and SARS.

To illustrate for adenovirus, the triplet distribution is the probability distribution $P_3(\cdot|A)$ on the 64-element set of all ordered triplets selected from the nucleotide alphabet $\{A,C,T,G\}$ given by
\begin{equation}
P_3(b_1b_2b_3|A) = \mathrm{Prob}\{A(K:[K+2]) = b_1b_2b_3\},
\label{eq.triplet-distribution}
\end{equation}
where $A(k:[k+2])$ is the length 3 substring of the adenovirus genome $A$ commencing at $k$ and $K$ is chosen at random from 1,$\dots$,34,123 (where the last triplet begins). Computation of $L_A(\cdot)$ from $P_3(\cdot|A)$ is accomplished by means of the analogous pair distribution $P_2(\cdot|A)$ and the second-order Markov transition matrix
\begin{eqnarray}
\lefteqn{T_3(b_1, b_2, b_3|A) = } \nonumber
\\
& & \mathrm{Prob}(A(k+2) = b_3| A(k) = b_1, A(k+1) = b_2).
\label{eq.transition-matrix}
\end{eqnarray}
The care necessary to handle Ns---present but undetermined bases---is exercised most easily with the pair distribution--transition matrix formulation.

In \cite{markovstructure-2021}, all three components of the Bayesian paradigm are varied, and their contributions to decisions resolved  The actual, error-containing reads can be replaced by error-free reads (available from the \MS), or degraded using the \MV, as in \cite{dqdegradation-2021}. The priors can be informative, as below, or uniform over $\mathcal{A}$, or even incorrect in the sense of omitting one genome \cite{markovstructure-2021}. Quality of the models can be reduced by degrading the reference genomes using the \MV.

The Bayesian analysis itself is straightforward: we use Bayes' theorem and the three likelihoods to calculate posterior probabilities over $\mathcal{A}$. Specifically, for $x \in \mathcal{A}$, the posterior probability of $x$ for read $R$ is
\begin{eqnarray}
\lefteqn{p(x|R) = } \nonumber
\\
& & \frac{\pi_R(x)L_x(R)}{\pi_R(A)L_A(R) + \pi_R(C)L_C(R) + \pi_R(S)L_S(R)}.
\label{eq.bayes}
\end{eqnarray}

\begin{figure}[h]
\centerline{
\includegraphics[width=3.4in]{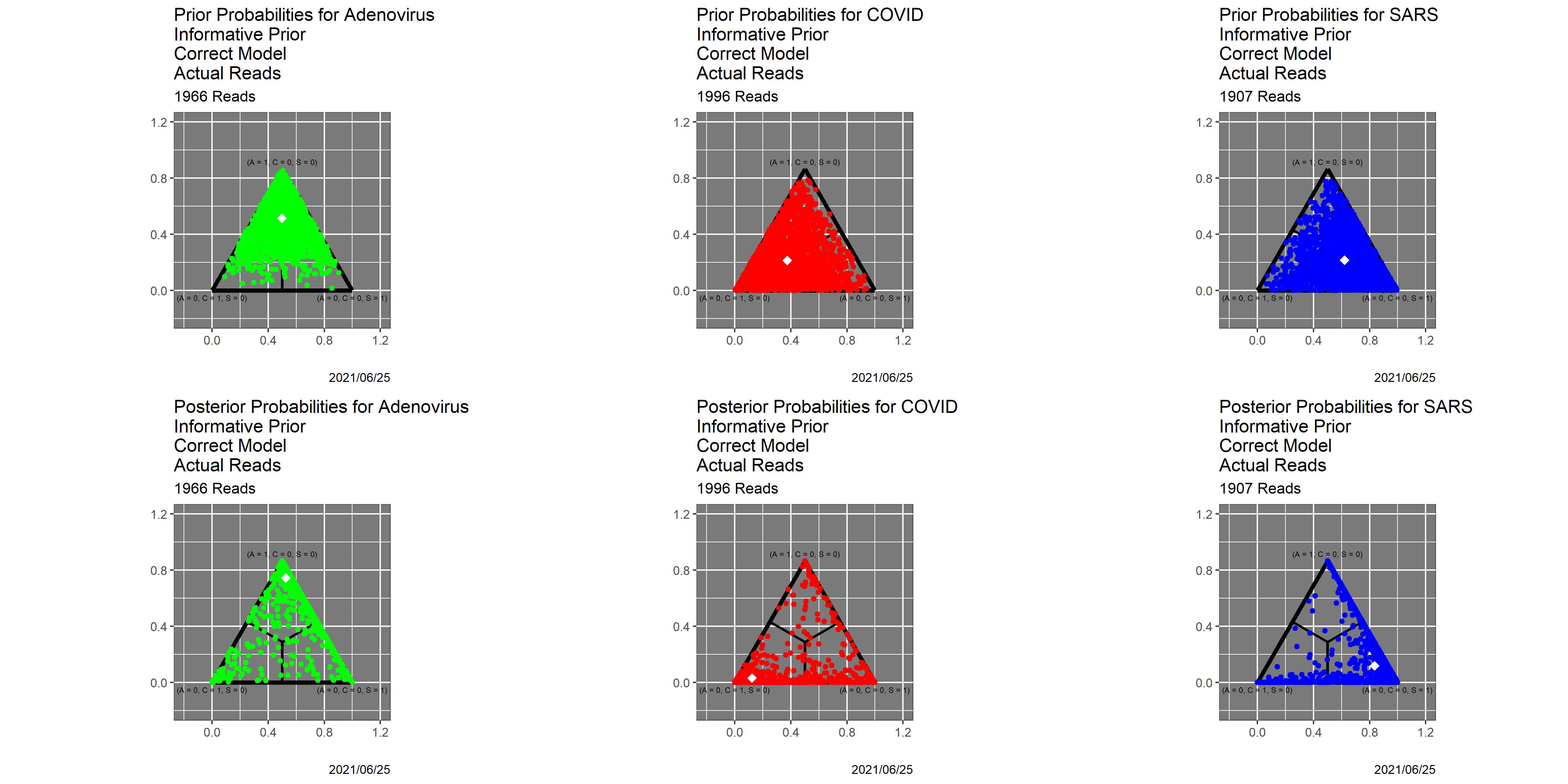}
}
\caption{Prior and posterior probabilities for the base case of actual reads, informative prior distributions and correct triplet distribution-based models. \emph{Top left:} prior probabilities for reads from adenovirus. \emph{Top middle:} prior probabilities for reads from COVID. \emph{Top right:} Prior probabilities for reads from SARS. \emph{Bottom left:} posterior probabilities for reads from adenovirus. \emph{Bottom middle:} posterior probabilities for reads from COVID. \emph{Bottom right:} Posterior probabilities for reads from SARS. White symbols in each plot are centroids.}
\label{fig.bayes-basecase}
\end{figure}

Figure \ref{fig.bayes-basecase} shows the prior and posterior probabilities for our base case of informative priors,\footnote{These were simulated using Dirichlet distributions that are biased modestly, albeit not uniformly, toward the truth.} the actual reads, and the correct models based on the triplet distribution likelihood functions from the three genomes. Because similar figures follow below, it is worthwhile to discuss it in some detail. First, there is one point for each read. Three-dimensional probabilities (barycentric coordinates) are represented in Cartesian coordinates as points in a two-dimensional simplex---an equilateral triangle. Pure adenovirus, in the sense that $\mathrm{Prob}(\mbox{Adeno}|R) = 1$, is the top vertex, pure COVID is the lower left vertex, and pure SARS is the lower right vertex. Because this is an experiment and we know the sources of the reads, there is a separate display for each source---adenovirus at the left, COVID in the center and SARS at the right. The upper three panels show prior probabilities, while the lower three panels show posterior probabilities. The green/red/blue coloring encoding read source is redundant but useful. The white dot in each graphic is the centroid of the probabilities it contains. The interior black lines, which are sometimes not visible, are the decision boundaries for the classifier that maximizes posterior probability over $\mathcal{A}$---the \emph{MAP classifier}.

Table \ref{tab.confusion-basecase} shows the confusion matrix for the MAP classifier. The performance is decent: the correct classification rate---calculable because this is an experiment with known ground truth---is 86.35\%.

\begin{table}
\begin{center}
\caption{Confusion matrix for the MAP classifier for the base case. Rows are read sources; columns are MAP decisions. The correct classification rate is 86.35\%.}
\begin{tabular}{|l|r|r|r|r|}
\hline
 & \multicolumn{3}{c|}{Decision} &
\\
\hline
Source & Adeno & COVID & SARS & Sum
\\
\hline
Adeno & 1757  &  74 & 135 & 1966
\\
\hline
COVID &   64 & 1762 & 170 & 1996
\\
\hline
SARS  &  214 &  144 & 1549 & 1907
\\
\hline
Sum  &  2035 & 1980 & 1854 & 5869
\\
\hline
\end{tabular}
\label{tab.confusion-basecase}
\end{center}
\end{table}

Many analyses would simply stop here, without reporting uncertainties, let alone making use of them. The bottom panel in Figure \ref{fig.bayes-basecase} demonstrates unequivocally that \emph{all decisions are not equal}, even when they are correct. In particular, for all three genomes, there are correctly classified reads that lie near the center of the triangle, meaning that they are close calls. To illustrate with an extreme case, if $p(\cdot|R) = (0.34, 0.33, 0.33)$, $R$ will be classified as adenovirus, even though many analysts might feel uncomfortable with this. Going one step further, the MAP classifier does not distinguish $R$ from $R'$ for which $p(\cdot|R') = (0.99, 0.005, 0.005)$.

\section{Specified Certainty Classification}\label{sec.scc}
\begin{figure}[h]
\centerline{
\includegraphics[width=2in]{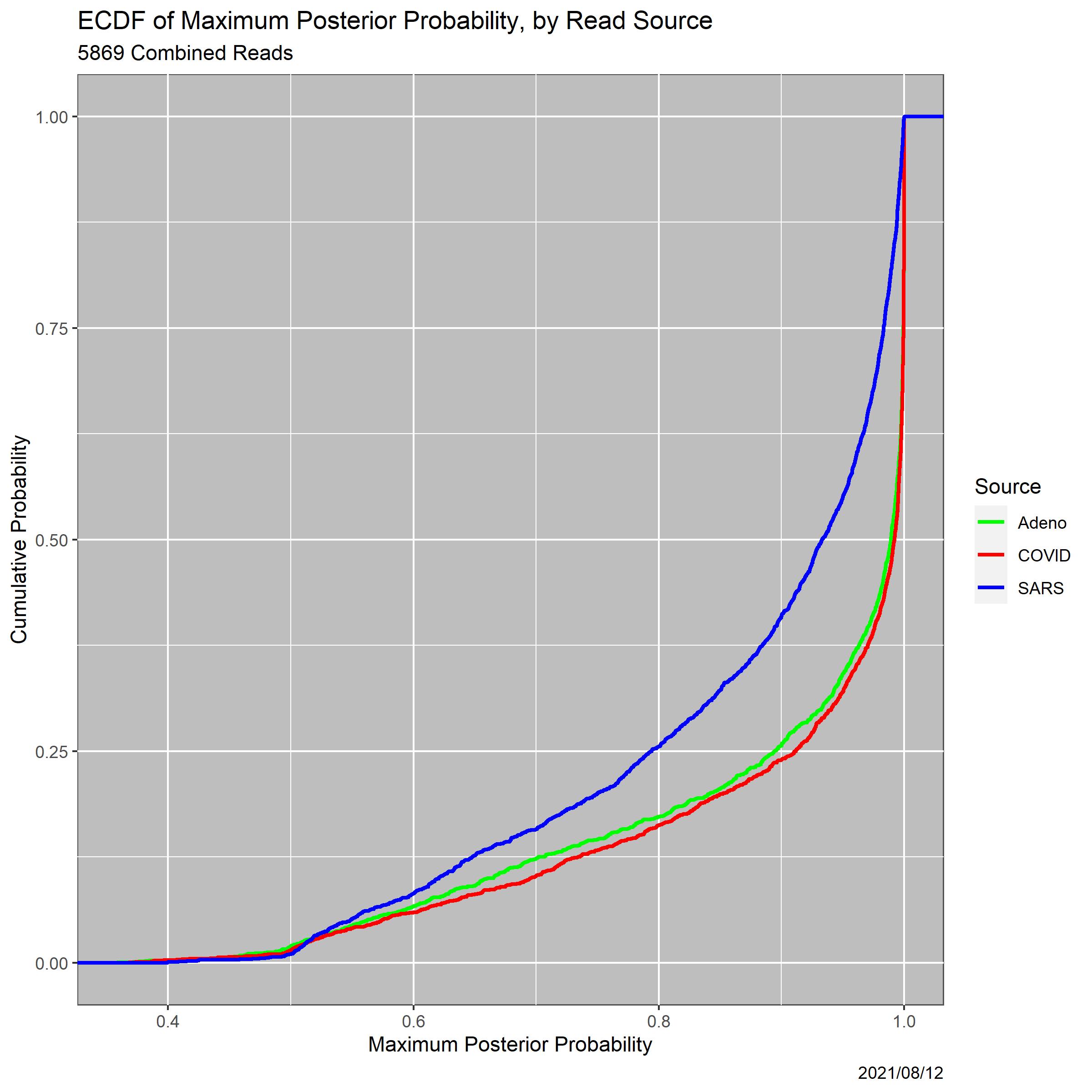}
}
\caption{ECDFs of certainty of MAP classifications, by read source.}
\label{fig.MAPvalue}
\end{figure}

One can go beyond common practice by at least reporting the uncertainties of the MAP classifications, for instance, summarizing them with the empirical cumulative distribution functions (ECDFs) in Figure \ref{fig.MAPvalue}. This figure establishes that many MAP decisions are not very certain. More than 15\% of all reads, and more than 25\% of the SARS reads, are classified with certainty less than 80\%. (Since the $p(\mathrm{Adeno}|R) + p(\mathrm{COVID}|R) + p(\mathrm{SARS}|R) = 1$ for each $R$, the maximum posterior probability cannot be less than 1/3.) In some circumstances, this may not be acceptable. Moreover, decisions for SARS are clearly, and statistically significantly, less certain than those for adenovirus or COVID. This distinction recurs throughout the paper.

\begin{figure}[h]
\centerline{
\includegraphics[width=2in]{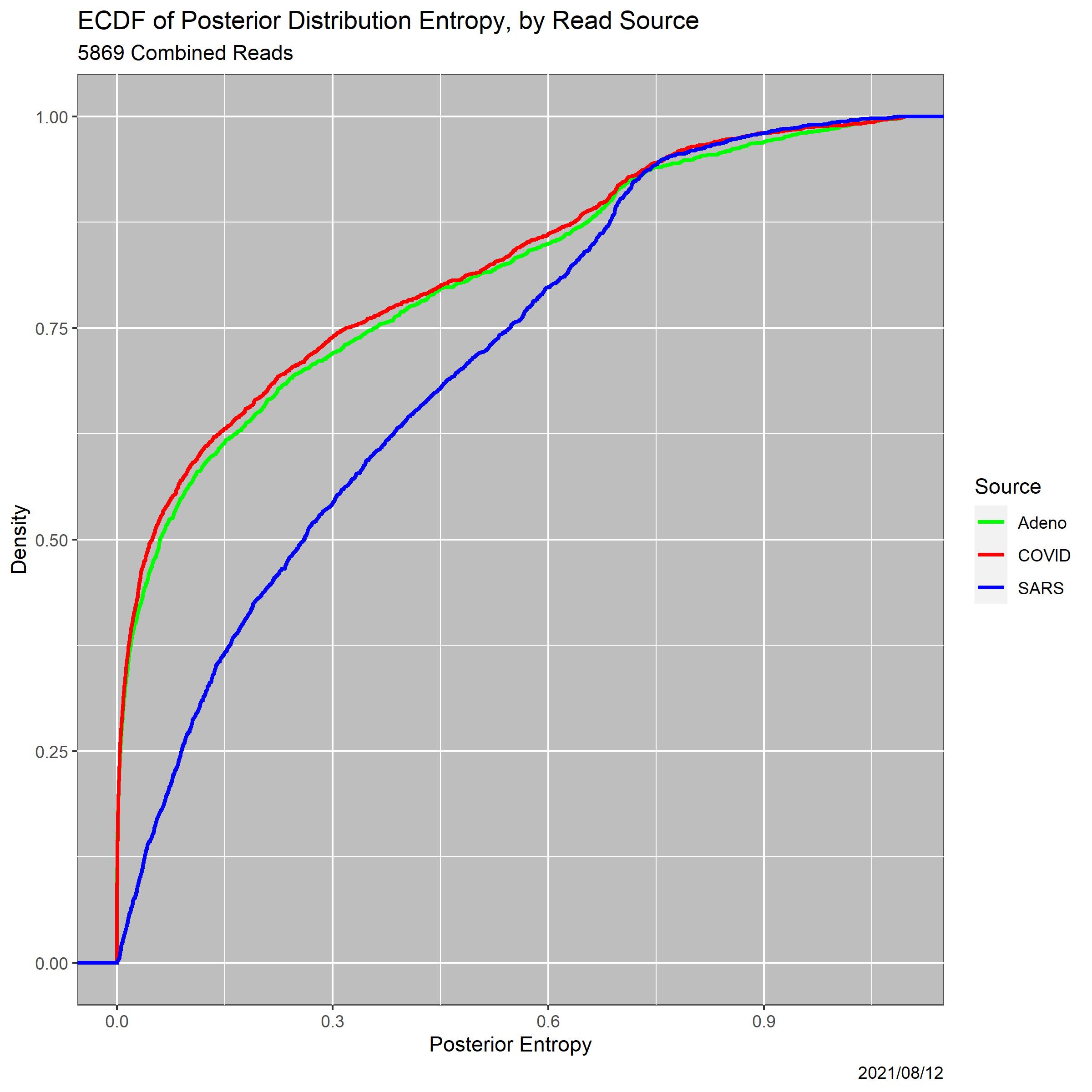}
}
\caption{ECDFs functions of posterior entropies, by read source.}
\label{fig.posteriorentropy}
\end{figure}

Complementarily, Figure \ref{fig.posteriorentropy} shows ECDFs of the entropies of the posterior distributions, again broken down by read source. The lower the entropy, the more definitive the MAP classification.\footnote{At the extreme, entropy is zero only for distributions concentrated at one point, and is maximized at the uniform distribution.} Three features of this figure are striking. First are the secondary modes, for all three sources, at approximately 0.7, which we do not pursue. The second is that the SARS ECDF differs dramatically from those for adenovirus and COVID, which do not differ significantly from one another. The two-sided $p$-values for Kolmogorov--Smirnov tests are 0 for COVID--SARS and 0.120 for COVID--adenovirus, Third and most important for this paper, for all three genomes there is considerable mass on values of entropy that do not represent clear-cut decisions by the MAP classifier.

\subsection{Formulation}\label{subsec.scc-formulation}
In response to these concerns, \SCC\ inverts and broadens the perspective. Instead of using MAP and having to live with the resultant uncertainties, the analyst specifies the desired (or required) level of certainty, and \SCC\ allows the classifier to produce less precise results that achieve it. Rather than forcing the classifier output to be one of 'Adeno', 'COVID' or 'SARS,' the atomic decisions, \SCC\ allows the classifier output to be \emph{any} nonempty subset of $\mathcal{A}$, as long as the certainty level exceeds the specified threshold. Mathematically, the allowable classifier outputs under \SCC\ are the power set of $\mathcal{A}$, excluding the empty set $\emptyset$:
\begin{eqnarray}
\mathcal{P}(\mathcal{A}) & = & \bigg\{\{\mbox{Adeno}\}, \{\mbox{COVID}\}, \{\mbox{SARS}\}, \label{eq.outputs}
\\
& & \{\mbox{Adeno, COVID}\}, \{\mbox{Adeno, SARS}\}, \nonumber
\\
& & \{\mbox{COVID, SARS}\}, \nonumber
\\
& & \{\mbox{Adeno, COVID, SARS}\} \bigg\}. \nonumber
\end{eqnarray}
Here lies one computational challenge: the number of elements in $\mathcal{P}(\mathcal{A})$ is $2^{|\mathcal{A}|}$, where $|S|$ is the cardinality of the set $S$. However, in applications, only the ``occupied'' elements of this set need to be enumerated or stored, which can be done using trees.

Computationally, \SCC\ is straightforward. Given the certainty level $\beta$, for each read $R$, sort the posterior probabilities $p(\mathrm{Adeno}|R)$, $p(\mathrm{COVID}|R)$ and $p(\mathrm{SARS}|R)$ in decreasing order, leading to a data structure
\begin{equation}
D =
\left[\begin{array}{ll}
A_1 & p(A_1|R)
\\
A_2 & p(A_2|R)
\\
A_3 & p(A_3|R)
\end{array}\right],
\label{eq.sccexample}
\end{equation}
where $(A_1, A_2, A_3)$ is a permutation of $\mathcal{A}$ and
$$
p(A_1|R) \geq p(A_2|R) \geq p(A_3|R).
$$
Let
$$
K = \min\big\{j:  p(A_1|R) + \dots  p(A_j|R) \geq \beta \big\};
$$
then the decision of $R$ is $\{A_1, \dots, A_K\}$. Note that $\beta = 1/ |\mathcal{A}|$ produces the MAP classifier, because $p(A_1|R) \geq \beta$ for all $R$.

With $\beta$ specified, therefore,
\begin{itemize}
\item
Some reads will be classified as `Adeno', as `COVID' or as `SARS,' meaning that with posterior certainty at least $\beta$ the read arises from exactly one of the three genomes.
\item
Other reads will be classified as `Adeno-or-COVID,' `Adeno-or-SARS', or `COVID-or-SARS,' meaning that with certainty $\beta$ they can only be said to have arisen from one of two genomes. This is the ``loss of precision'' referred to in \S\ref{sec.introduction}.
\item
Still other reads will be classified as `Adeno-or-COVID-or-SARS,' the equivalent of ``unclassifiable.''
\end{itemize}
Below, we refer to these as decisions of Precision 1, 2 and 3, respectively.

\begin{figure}[h]
\centerline{
\includegraphics[width=3.4in]{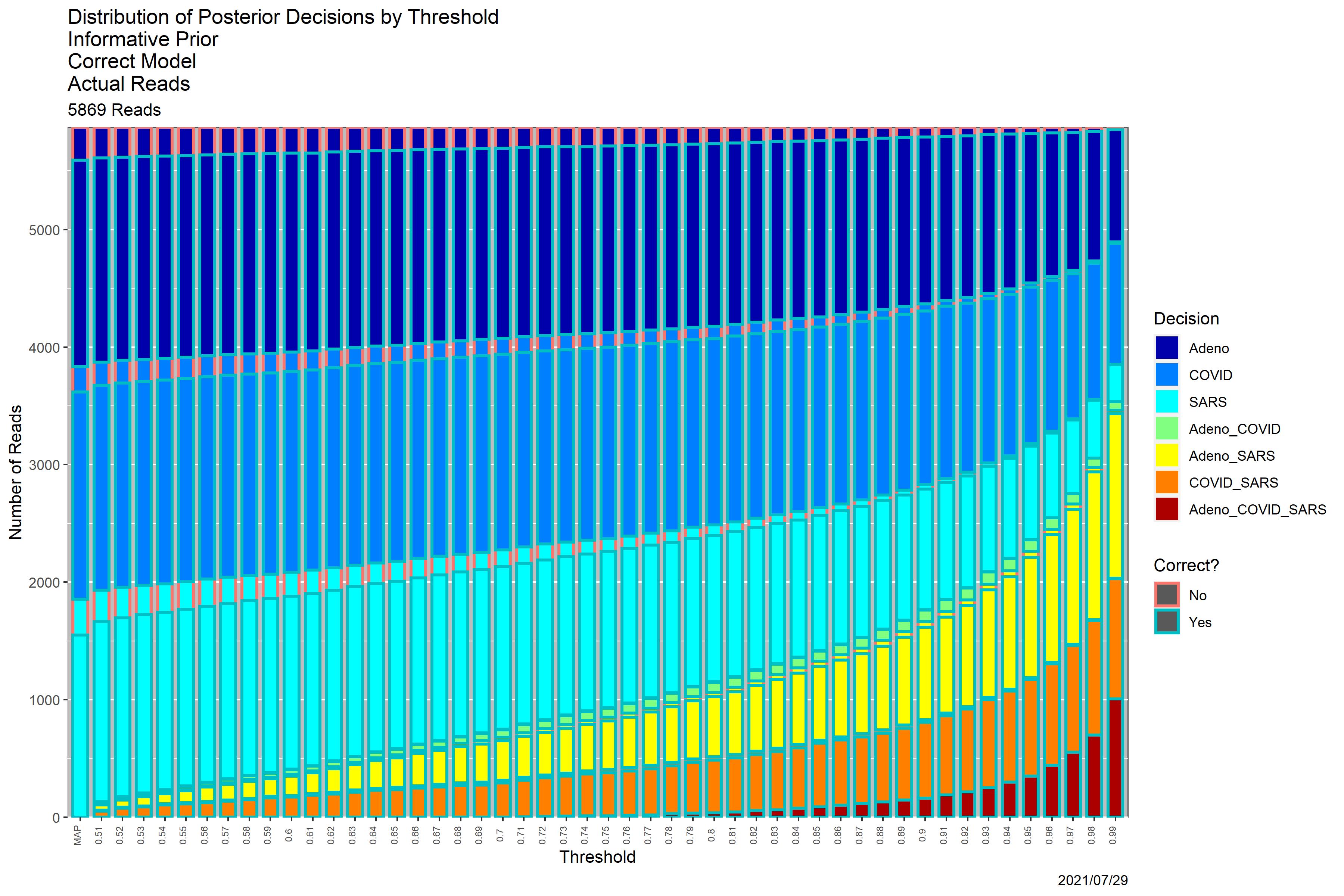}
}
\caption{\SCC\ output for the posterior probabilities in Figure \ref{fig.bayes-basecase}.}
\label{fig.scc-basecase-posterior}
\end{figure}

Figure \ref{fig.scc-basecase-posterior}, which is in some sense our ``takeaway message,'' shows the result of applying \SCC\ to the reads dataset, using the posterior probabilities appearing in Figure \ref{fig.bayes-basecase}. The horizontal axis contains thresholds of 0.33 (as noted above, the MAP estimator) and $0.51, \dots, 0.99$. The bar for each threshold shows the distribution of classifier output for that threshold, split in two ways. Color of the bar encodes the decision, and color of the surrounding box indicates whether the decision is correct. ``Correct'' no longer means an exact match. For instance, if the source of read $R$ is adenovirus, the decision is correct if the classifier output for $R$ is any one of
`Adeno', `Adeno-or-COVID,' `Adeno-or-SARS,' or `Adeno-or-COVID-or-SARS.' While ground truth is required to determine correctness (box color), it is not required to determine the distribution of decisions (bar color).

\subsection{Tradeoffs}\label{subsec.tradeoffs}
Qualitatively, Figure \ref{fig.scc-basecase-posterior} shows exactly the behavior we expected. As the threshold increases, precision decreases. For low thresholds (at the left), Precision 1 (a single genome) decisions predominate, and while some of these remain for high thresholds (at the right) most decisions become Precision 2 or 3. The same can be seen more quantitatively by comparing the ''Posterior Probabilities'' columns in Tables \ref{tab.threshold80} ($\beta = 0.80$) and \ref{tab.threshold99} ($\beta = 0.99$) below. There is no universally correct choice of threshold. Indeed, \SCC, as in \cite{recordlinkage-plos1-2019}, is to empower the user by illuminating the consequences of using it.

\begin{figure}[h]
\centerline{
\includegraphics[width=2in]{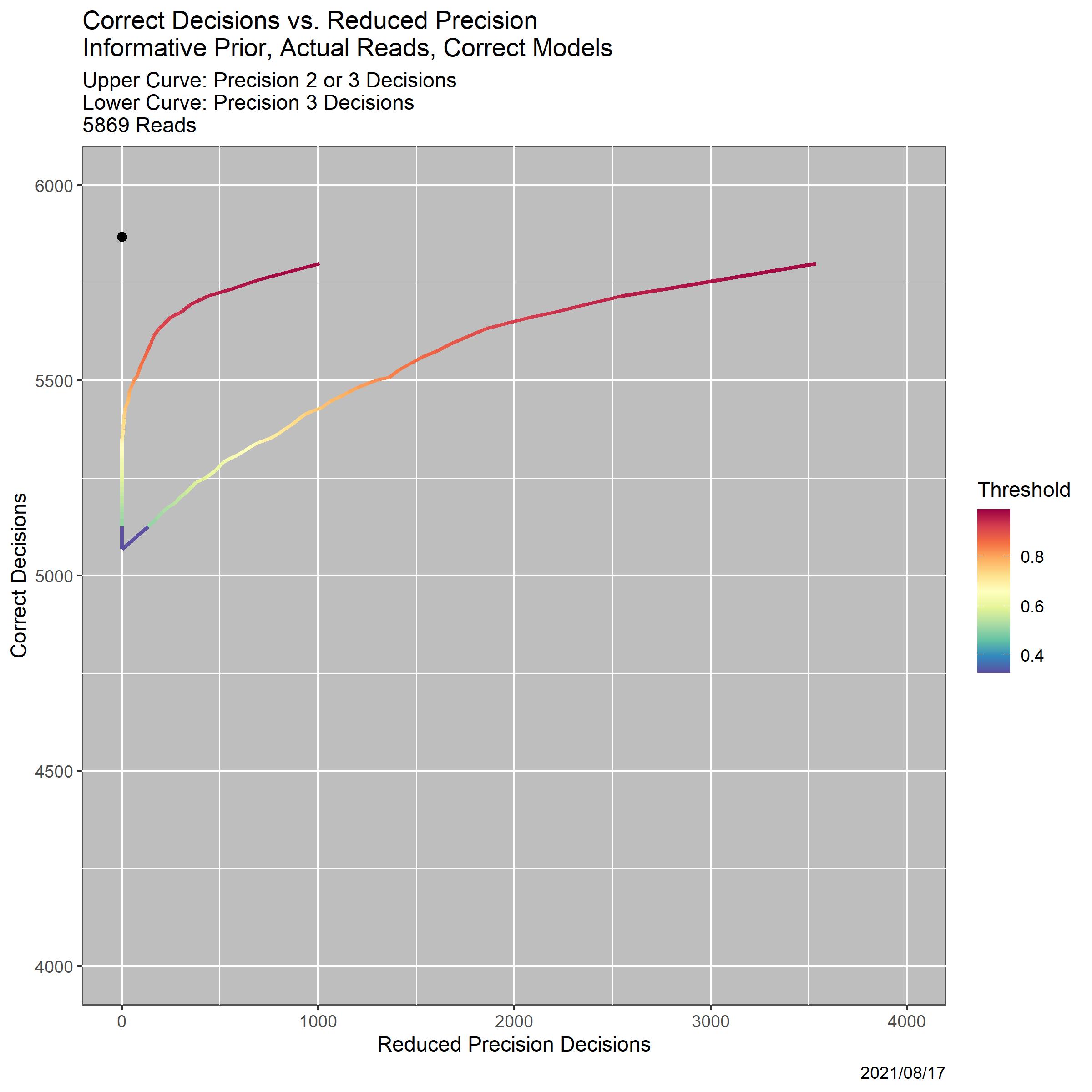}
}
\caption{Tradeoff of correctness and precision for the example in Figure \ref{fig.scc-basecase-posterior}. \emph{Lower curve:} incorrect decisions vs. Precision 2 decisions. \emph{Upper curve:} Incorrect decisions vs. Precision 2 or 3 decisions.}
\label{fig.scc-precisionvscorrect}
\end{figure}

There are also more subtle behaviors. Figure \ref{fig.scc-precisionvscorrect} shows how \SCC\ trades off precision and correctness. There are two curves, each parameterized by the threshold $\beta$. The $y$-axis is always the number of correct reads. The $x$-axis is the number of ``less precise reads,'' which is different for the two curves, and explained momentarily. As for \ROC\ curves, optimality is the black dot at the upper left, representing 5869 correct reads and no less precise reads. For the upper curve, less precise means Precision 2 or 3. For the lower curve, it means Precision 3 alone.

\subsection{Classifier Behavior}\label{subsec.scc-behavior}
\begin{figure}[h]
\centerline{
\includegraphics[width=3.4in]{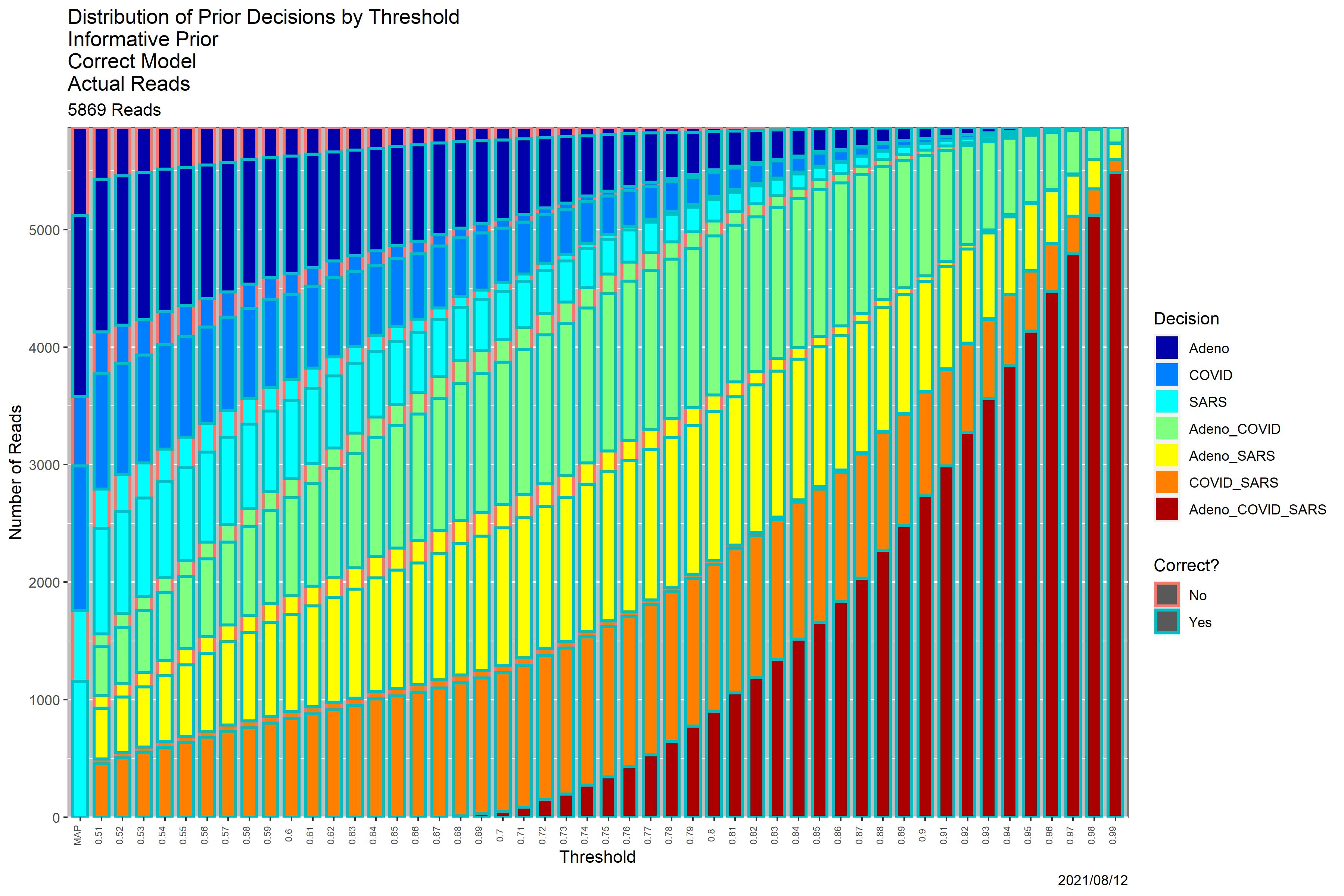}
}
\caption{\SCC\ output for the prior probabilities in Figure \ref{fig.bayes-basecase}.}
\label{fig.scc-basecase-prior}
\end{figure}

\SCC\ also illuminates the classifier itself. First, \SCC\ can be applied to prior probabilities. Figure \ref{fig.scc-basecase-prior} contains \SCC\ results for the prior probabilities for the base case. The Bayesian computation in (\ref{eq.bayes}) transforms the decision structure in Figure \ref{fig.scc-basecase-prior} into that in Figure \ref{fig.scc-basecase-posterior}. Uniformly for all thresholds, using posterior probabilities leads to larger numbers of high precision decisions than using the priors alone. At an extreme, for $\beta = 0.99$ (the far right in both figures) using prior probabilities, nearly all reads are classifiable only as `Adeno-or-COVID-or-SARS,' whereas when posterior probabilities are employed, only approximately 15\% are. Table \ref{tab.threshold99} provides a detailed comparison. This table quantifies how dramatically decisions are improved by adding data and models to the prior probabilities. Even for the much less conservative threshold of $\beta = 0.80$, similar behavior obtains, as shown in Table \ref{tab.threshold80}.

\begin{table}
\begin{center}
\caption{\SCC\ results for threshold $\beta = 0.99$ using prior and posterior probabilities that appear in Figure \ref{fig.bayes-basecase}.}
\begin{tabular}{lrr}
\hline
& \multicolumn{2}{c}{Probabilities}
\\
Decision & Prior & Posterior
\\
Adeno & 0 & 975
\\
COVID & 1 & 1042
\\
SARS & 1 & 316
\\
Adeno-or-COVID & 131 & 73
\\
Adeno-or-SARS & 139 & 1006
\\
COVID-or-SARS & 111 & 1025
\\
Adeno-or-COVID-or-SARS & 5486 & 1075
\\
\hline
\end{tabular}
\label{tab.threshold99}
\end{center}
\end{table}

\begin{table}
\begin{center}
\caption{\SCC\ results for threshold $\beta = 0.80$ using prior and posterior probabilities that appear in Figure \ref{fig.bayes-basecase}.}
\begin{tabular}{lrr}
\hline
& \multicolumn{2}{c}{Probabilities}
\\
Decision & Prior & Posterior
\\
\hline
Adeno & 363 & 1694
\\
COVID & 227 & 1689
\\
SARS & 208 & 1334
\\
Adeno-or-COVID & 1479 & 80
\\
Adeno-or-SARS & 1409 & 548
\\
COVID-or-SARS & 1280 & 475
\\
Adeno-or-COVID-or-SARS & 903 & 39
\\
\hline
\end{tabular}
\label{tab.threshold80}
\end{center}
\end{table}

\begin{figure}[h]
\centerline{
\includegraphics[width=3.4in]{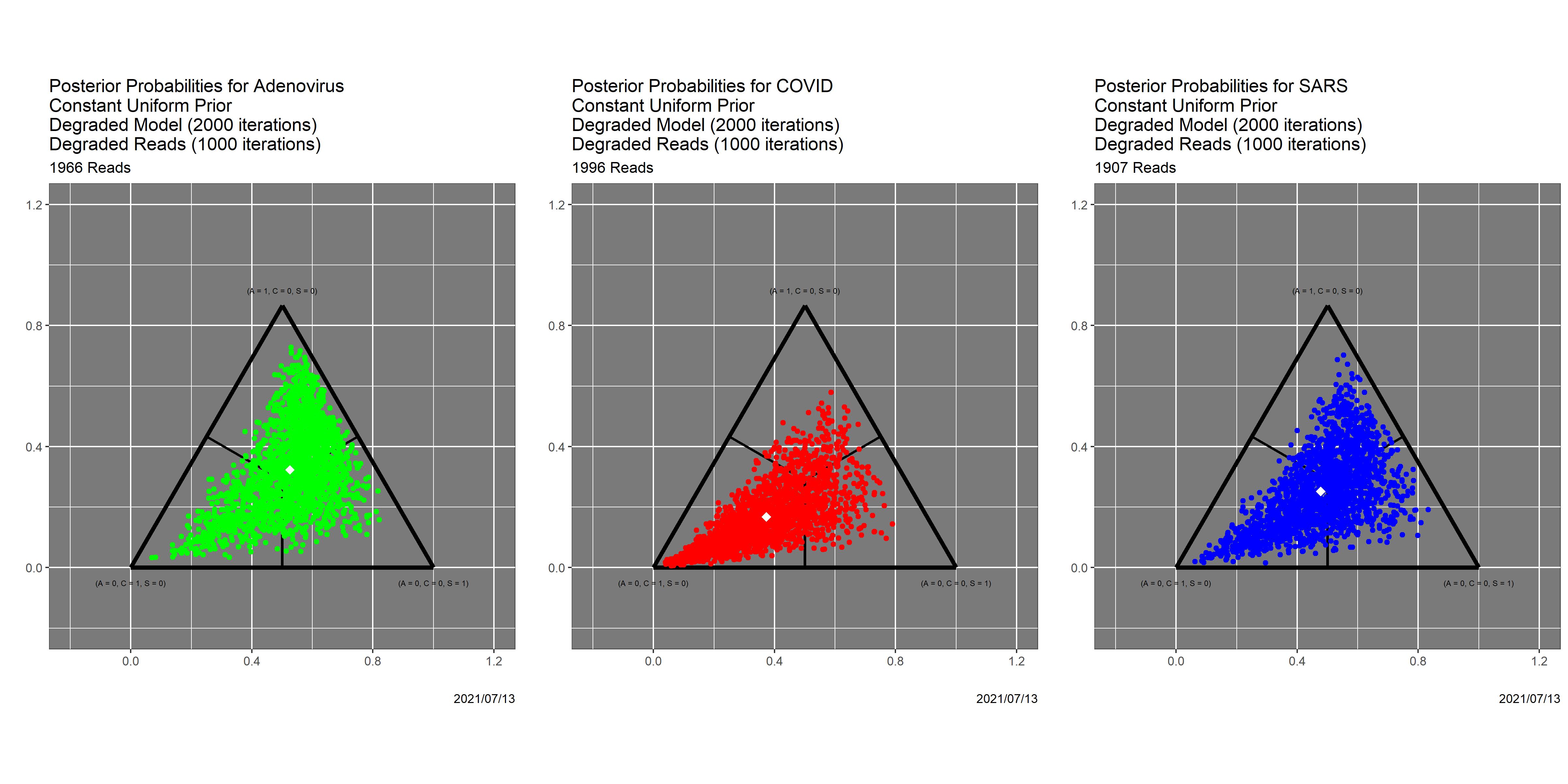}
}
\caption{Posterior probabilities for the ``worst of all possible worlds'' scenario.}
\label{fig.bayes-worstcase}
\end{figure}

In \cite{markovstructure-2021} we explore in depth the effects of varying the prior, the quality of the read data, and the quality of the models, which reflects the quality of the training data. Without replicating the granularity of that exploration, we show here how \SCC\ delivers complementary insights. The ``worst of all possible worlds'' scenario in \cite{markovstructure-2021} is one in which all priors are uniform over $\mathcal{A}$, reads are degraded by 1000 iterations of the \MV\ and the training data underlying the likelihood functions are degraded by 2000 iterations of the \MV\ on the actual genomes, before calculating the triplet distributions and likelihood functions.. The posterior probabilities appear in Figure \ref{fig.bayes-worstcase}. Only the most cursory visual comparison with Figure \ref{fig.bayes-basecase} is needed to know that something is seriously wrong. Indeed, this figure more closely resembles Figure \ref{fig.scc-basecase-prior} than Figure \ref{fig.bayes-basecase}. In other words, the starting point for the base case does not differ dramatically from the ending point for the ``worst of all possible worlds'' scenario.

\begin{figure}[h]
\centerline{
\includegraphics[width=3.4in]{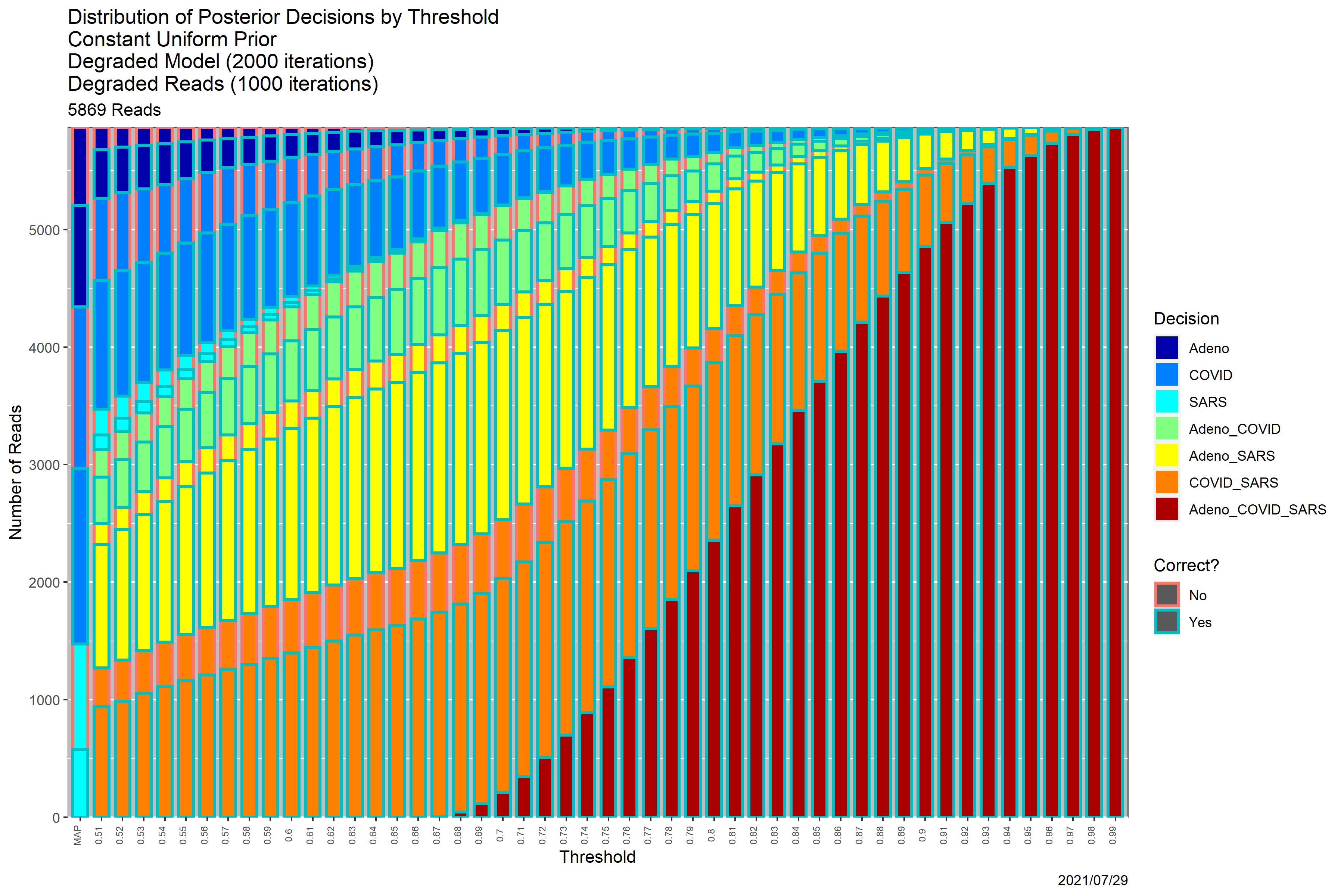}
}
\caption{\SCC\ output for the ``worst of all possible worlds'' posterior probabilities in Figure \ref{fig.bayes-worstcase}.}
\label{fig.scc-worstcase-posterior}
\end{figure}

Figure \ref{fig.scc-worstcase-posterior} shows that the \SCC\ consequences are similarly deleterious. Again choosing $\beta = 0.80$ as exemplar, Table \ref{tab.threshold80-worstcase} contains the associated decisions; it should be compared to the ``Posterior Probabilities'' column of Table \ref{tab.threshold80}.

\begin{table}
\begin{center}
\caption{\SCC\ results for threshold $\beta = 0.80$ and the ``worst of all possible worlds'' scenario.}
\begin{tabular}{lr}
\hline
Decision & Number of Reads
\\
\hline
Adeno &  9
\\
COVID & 205
\\
SARS & 0
\\
Adeno-or-COVID & 330
\\
Adeno-or-SARS & 1166
\\
COVID-or-SARS & 1803
\\
Adeno-or-COVID-or-SARS & 2356
\\
\hline
\end{tabular}
\label{tab.threshold80-worstcase}
\end{center}
\end{table}

\subsection{Transition Points}\label{subsec.scc-transition}
\begin{figure}[h]
\centerline{
\includegraphics[width=2in]{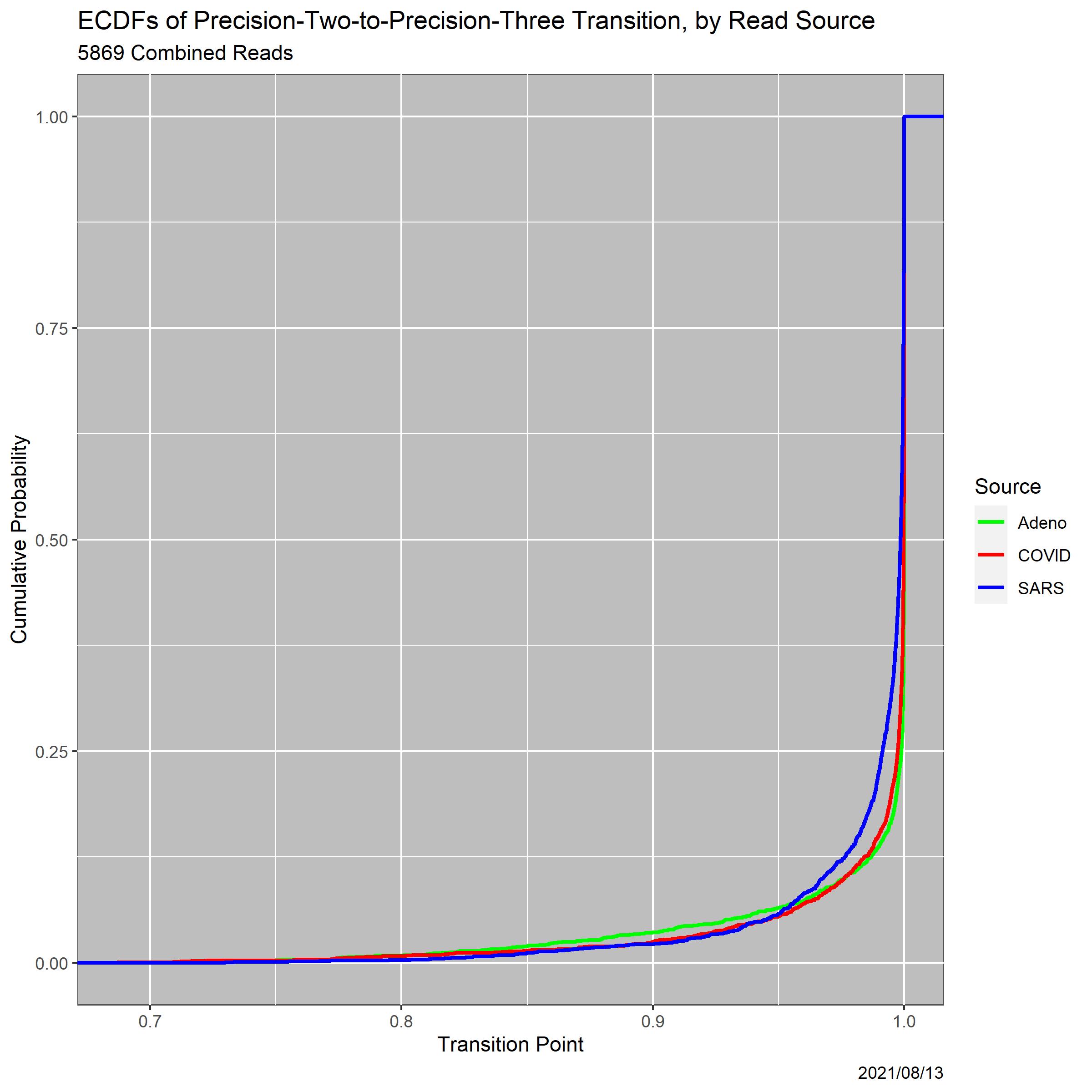}
}
\caption{ECDFs of Precision-2-to-Precision-3 transitions, by read source.}
\label{fig.transition2to3}
\end{figure}

A third pathway to insight, which arose in the example in \S\ref{sec.alternatives} (see Table \ref{tab.transitions}), is to examine the thresholds at which loss of precision occurs. Each read undergoes two loss-of-precision transitions, for instance, from `Adeno' to `Adeno-or-COVID' and then to `Adeno-or-COVID-or-SARS.' ECDFs of the transitions are informative. Since the Precision-1-to-Precision-2 transition is identical to the certainty of the MAP estimator, Figure \ref{fig.MAPvalue} contains the ECDFs of the Precision-1-to-Precision-2 transitions. Figure \ref{fig.transition2to3}, is analogous, but for Precision-2-to-Precision-3 transitions. Notably, the SARS versus Adeno and COVID gap for Precision-1-to-2 transitions is diminished significantly, although it does not vanish and remains statistically significant.

\subsection{Collective Certainty}\label{subsec.scc-multiplicity}
This material is narrow but important. The certainty levels promised by \SCC\ apply to each data point (read, in this case) individually, not to the collective of decisions. There is an extensive literature on multiple testing in statistics \cite{westfallyoung93, westfall-mc-2016}. Whether \SCC\ decisions can be treated in the same way as the results of hypothesis tests is not clear at this point. Bonferroni corrections, meaning essentially that the decisions are treated as independent, may lead extremely small estimates of collective certainty, especially for large datasets. Alternatives such as false discovery rates \cite{benjaminihochberg-1995} may be more promising. An initial example appears in \S\ref{sec.alternatives}, where the effect of increasing the threshold is unmistakeable.

\section{Application to COVID-19 Vaccination Data}\label{sec.alternatives}
The \SCC\ paradigm works in any context where the classifier output consists of (posterior) probabilities on a finite set $\mathcal{A}$ of atomic decisions. To illustrate, in this section we analyze state-level COVID-19 vaccination rates as they relate to the outcome of the 2020 election (Democratic or Republican, again by state). The analysis data consist of 51 state proportions $V(S)$ of the entire population who were fully vaccinated as of July 31, 2021, which were downloaded from the \CDC\ website.\footnote{The vaccination data, updated weekly, are at https://covid.cdc.gov/covid-data-tracker/\#vaccinations. Weekly state-level case and death counts are also available.} To give a sense of the data, the fully-vaccinated proportions vary from 34.3\% (AL) to 67.5\% (VT).

\subsection{Mixture Model}\label{subsec.alternatives-mixture}
\begin{figure}[h]
\centerline{
\includegraphics[width=1.5in]{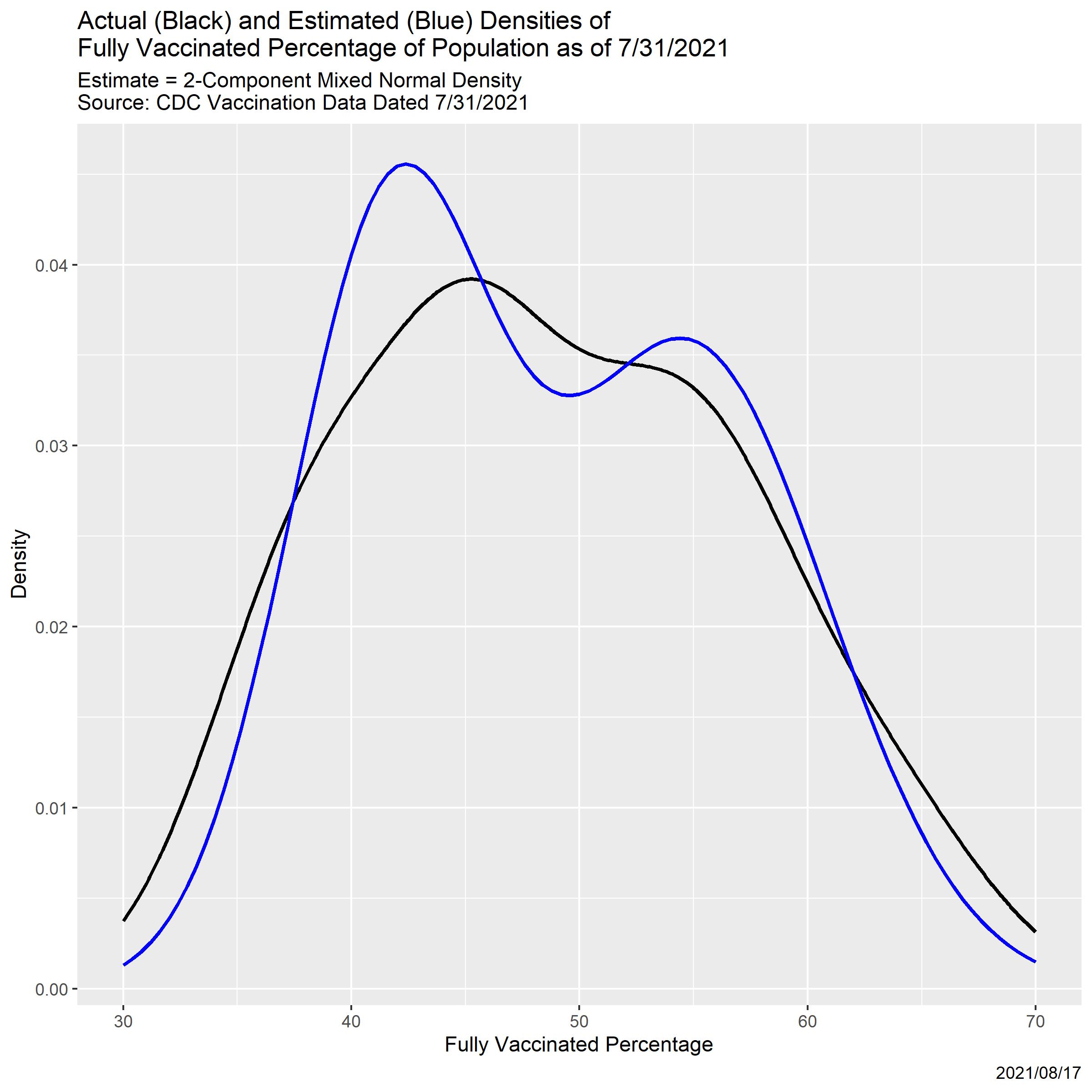}
}
\caption{Actual and estimated densities of fully COVID-19 vaccinated percentages as of 7/31/2021.}
\label{fig.vax-density}
\end{figure}

Figure \ref{fig.vax-density} shows in black the density function of the 51 fully vaccinated percentages. (The blue curve will be explained momentarily.) There is clear evidence of bimodality, so it is natural to model this density as a two-component mixture of normal distributions. We did so using the \texttt{mixtools} package in R \cite{mixtools-2009}, whose algorithm derives from \cite{mclachlan-mixture-2000}. Table \ref{tab.mixturecomponents} shows the weights, means and standard deviations of the two components of the resulting mixture model. The blue line in Figure \ref{fig.vax-density} is the fitted density. The fit is better than Figure \ref{fig.vax-density} may seem to suggest: of 1000 bootstrapped samples of size 51 from the mixed normal density, only three led to rejection of the hypothesis that the two densities are identical.

\begin{table}[h]
\begin{center}
\caption{Parameters of the Mixed Normal Model.}
\begin{tabular}{lrr}
\hline
Parameter & Component 1 & Component 2
\\
\hline
Weight $\lambda$ & 0.4678343 & 0.5321657
\\
Mean $\mu$ & 41.7228 & 54.8779
\\
Standard deviation $\sigma$ & 4.447488 & 6.008174
\\
\hline
\end{tabular}
\label{tab.mixturecomponents}
\end{center}
\end{table}


Often mixing arises from an unobserved, or at least unmodeled, \emph{latent variable} \cite{bollen-lv-1989} taking as many possible values as there are mixture components---in this case, two. SPOILER ALERT: as the colors in Figure \ref{fig.scc-vax} suggest, we hypothesize that the latent variable measures the outcome of the 2020 Presidential election, which we denote by $O(S) \in \{D, R\}$. Note that this interpretation of the mixture model is consistent with modern thinking regarding causality \cite{bookofwhy-2018}, in the sense that the potential causality is in the right direction.\footnote{An alternative approach that estimated two densities for vaccination, one for states with $O(S) = D$ and one for states with $O(S) = R$, would at the person level be highly suspect because it posits that vaccination causes voting behavior. Not only is that relationship retroactive in time, but also there is no scientifically plausible causal path.} The relationship might still, of course, not be causal: voting and vaccination could both result from common underlying causal factors such as age, education, income or race.

\subsection{Application of SCC}\label{subsec.alternatives-scc}
\begin{figure}[h]
\centerline{
\includegraphics[width=2in]{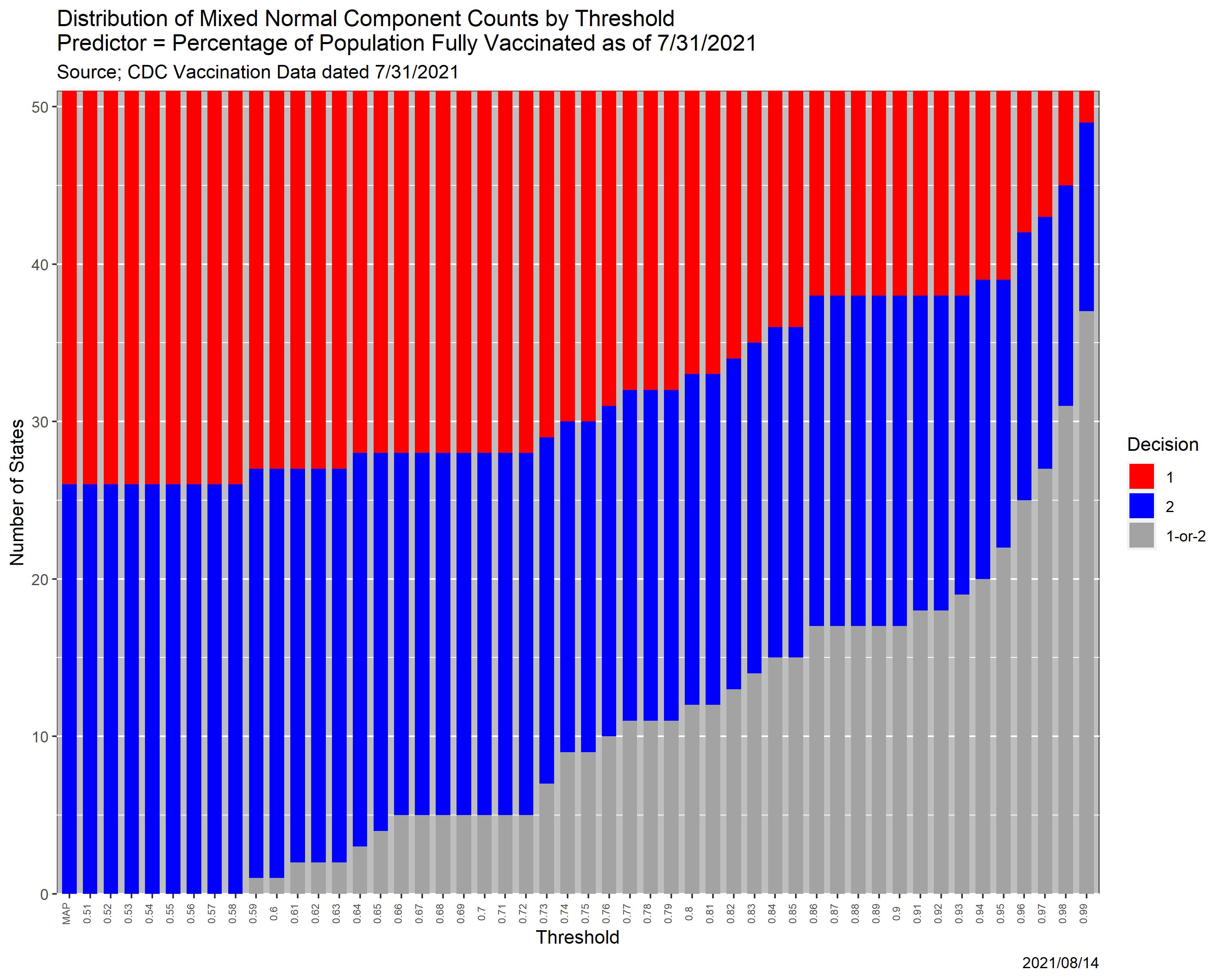}
}
\caption{\SCC\ results for 2-component mixed normal model of state-level percentages of fully COVID-vaccinated population.}
\label{fig.scc-vax}
\end{figure}

Returning to the mixture model, reflecting the unobserved nature of latent variables, many fitting methods, including \texttt{mixtools}, employ the iterative EM (expectation-maximization) algorithm \cite{dlr-em77}. The primary outputs, in consequence, are posterior distributions over the components for each data point. Mixture models are, therefore, natural candidates for \SCC, and Figure \ref{fig.scc-vax} contains the results of applying it to the case at hand.

The qualitative interpretations are similar to those in \S\ref{sec.scc}. With two outcomes, the MAP estimator corresponds to $\beta = 1/2$. Figure \ref{fig.scc-vax}---more precisely, the numerical values underlying it---is more informative scientifically that any single set of predictions. For instance, contrast with \S\ref{sec.scc}, there are states that are predicted to be a single component even when $\beta = 0.99$. Two states, AL and MS, are always in Component 1, while CT, MA, MD, ME, NH, NJ, NM, NY, OR, RI, VT, and WA (12 states) are always in Component 2. Only once $\beta \geq .59$ is any state not placed in a single component; from Table \ref{tab.transitions}, that state is SD.

Figure \ref{fig.scc-vax} does not,\emph{ per se}, dictate what threshold should be chosen in a given context. It does allow an analyst to understand his or her choice of a threshold in light of alternatives. To illustrate, there is no reason to choose $\beta \in [.66, .71]$ because the results are identical to those for $\beta = .72$. Figure \ref{fig.scc-vax} does not even dictate that a single threshold must be chosen, because considering all thresholds is more informative.

As intimated, the colors in Figure \ref{fig.scc-vax} were selected deliberately. Table \ref{tab.component-vs-election} is a cross-tabulation of state-level results in the 2020 Presidential election with the MAP component assignments. Only 6 states are ``mis-classified:'' AZ, GA, and NV voted Democratic but are in Component 1, while FL, IA and NE voted Republican but are in Component 2.\footnote{Other analyses not reported here show these six states to be ``vaccination outliers.'' For instance, the high population of seniors in FL, most of whom are vaccinated, distorts an otherwise low vaccination rate.} Illustrating the kinds of insights \SCC\ facilitates, Table \ref{tab.component-vs-election-80-98} shows the same information for $\beta = 0.80$ and $0.98$, the minimum threshold for which there are no errors.

\begin{table}
\begin{center}
\caption{Cross-tabulation of MAP component assignments and state-level 2020 Presidential election outcomes.}
\begin{tabular}{lrr|r}
\hline
& \multicolumn{2}{c}{2020 Election Outcome}
\\
Component & Democratic & Republican & Sum
\\
\hline
1 &   3 & 22 &  25
\\
2 &  23 & 3 & 26
\\
1-or-2 &  0 & 0 &  0
\\
\hline
Sum & 26 & 25 & 51
\\
\hline
\end{tabular}
\label{tab.component-vs-election}
\end{center}
\end{table}

\begin{table}
\begin{center}
\caption{\emph{Top:} Cross-tabulation of component assignments for $\beta = 0.80$ and state-level 2020 Presidential election outcomes. \emph{Bottom:} Cross-tabulation of component assignments for $\beta = 0.98$ and state-level 2020 Presidential election outcomes.}
\begin{tabular}{lrr|r}
\hline
& \multicolumn{2}{c}{2020 Election Outcome}
\\
Component & Democratic & Republican & Sum
\\
\hline
1 &   2 & 16 &  18
\\
2 &  21 & 0 & 21
\\
1-or-2 &  2 & 9 & 12
\\
\hline
Sum & 26 & 25 & 51
\\
\hline
\end{tabular}

\vspace{.2in}
\begin{tabular}{lrr|r}
\hline
& \multicolumn{2}{c}{2020 Election Outcome}
\\
Component & Democratic & Republican & Sum
\\
\hline
1 &   0 & 6 &  6
\\
2 &  14 & 0 & 14
\\
1-or-2 &  12 & 19 & 31
\\
\hline
Sum & 26 & 25 & 51
\\
\hline
\end{tabular}
\label{tab.component-vs-election-80-98}
\end{center}
\end{table}

Because there are only 51 data points and only one loss-of-precision transition for each, loss-of-precision for the analysis in Figure \ref{fig.scc-vax} can be enumerated, resulting in Table \ref{tab.transitions}. Each state undergoes one transition as $\beta$ increases, from either `1' or `2' to `1-or-2;' these are the values in the table.

\begin{table}
\begin{center}
\caption{Transition points associated with Figure \ref{fig.scc-vax}.}
\begin{tabular}{llr|llr}
  \hline
State & 2020 & Loss of & State & 2020 & Loss of
\\
& Outcome & Precision & & Outcome & Precision
\\
  \hline
  AK & R & 73.68 & MT & R & 82.55 \\
  AL & R & 99.05 & NC & R & 85.35 \\
  AR & R & 98.50 & ND & R & 95.81 \\
  AZ & D & 75.38 & NE & R & 72.24 \\
  CA & D & 94.63 & NH & D & 99.87 \\
  CO & D & 97.99 & NJ & D & 99.88 \\
  CT & D & 100.00 & NM & D & 99.70 \\
  DC & D & 98.45 & NV & D & 81.93 \\
  DE & D & 94.31 & NY & D & 99.68 \\
  FL & R & 65.37 & OH & R & 63.81 \\
  GA & D & 97.25 & OK & R & 95.54 \\
  HI & D & 96.47 & OR & D & 99.31 \\
  IA & R & 73.32 & PA & D & 92.85 \\
  ID & R & 98.07 & RI & D & 99.99 \\
  IL & D & 60.48 & SC & R & 95.09 \\
  IN & R & 83.15 & SD & R & 58.12 \\
  KS & R & 76.20 & TN & R & 96.91 \\
  KY & R & 72.79 & TX & R & 85.35 \\
  LA & R & 98.31 & UT & R & 79.24 \\
  MA & D & 100.00 & VA & D & 98.12 \\
  MD & D & 99.91 & VT & D & 100.00 \\
  ME & D & 100.00 & WA & D & 99.78 \\
  MI & D & 64.16 & WI & D & 90.58 \\
  MN & D & 97.07 & WV & R & 97.00 \\
  MO & R & 93.62 & WY & R & 98.39 \\
  MS & R & 99.01 &  &  &  \\
   \hline
\end{tabular}
\label{tab.transitions}
\end{center}
\end{table}

\begin{figure}[h]
\centerline{
\includegraphics[width=2in]{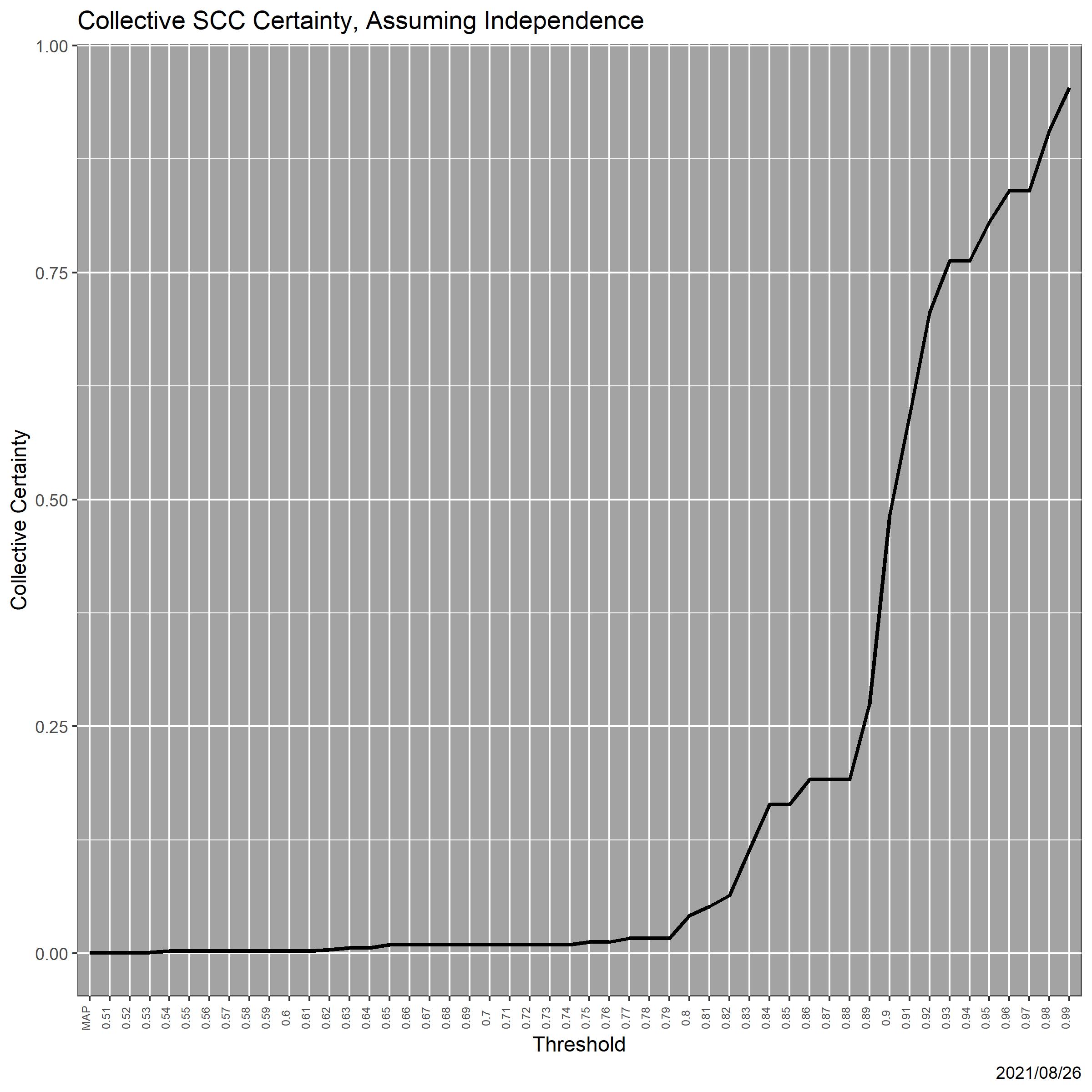}
}
\caption{Collective certainty as a function of threshold.}
\label{fig.scc-collectivecertainty}
\end{figure}

Finally, returning to collective certainty (\S\ref{subsec.scc-multiplicity}), Figure \ref{fig.scc-collectivecertainty} shows the dramatic increase in collective certainty relative under the heuristic and \emph{ad hoc} scenario of independence of decisions over states.

\section{Conclusion}\label{sec.conclusion}
\SCC\ empowers analysts and decision makers. First and foremost, it creates control of the level of certainty in classifier output by allowing some decisions to be less precise than others. Contrastingly, MAP classifiers classify every data point with Precision 1; \SCC\ shows the potentially negative implications of doing so.  \SCC\ also provides unprecedented insight into the behavior of classifiers and the consequences with respect to making decisions. Moreover, \SCC\ allows those of us who develop tools to be scientifically agnostic about their use: \SCC\ does not make decisions, but instead provides an explainable basis for making principled decisions. Finally, \SCC\ is a step in the direction of interpretability, a major issue for \AI\ \cite{interpretibleAI-2021}.

Much remains to be done. Many classifiers, especially {\AI}-based deep learning models and trees, do not currently provide the posterior probabilities required by \SCC. One potential path is that warned against in \cite{interpretibleAI-2021}, namely developing approximations that do produce uncertainties.\footnote{An intriguing analogy is computer experiments \cite{sacks-etal-computerexperiments-1989, santner-computerexperiments-2003}, in which Gaussian process priors are used to develop approximations to complex numerical models.}

Extension is also needed to cases where our ``allow all subsets of the space of atomic decisions'' is not feasible.\footnote{For example, if the atomic decisions are ages, the feasible compound decisions might be constrained to be age ranges, because a decision ``age = 23-or-27-or-31'' is nonsensical.} The necessary infrastructure is that the set of allowable compound decisions must be partially ordered by set inclusion.

\bibliographystyle{apalike}
\bibliography{AFK}

\begin{thebibliography}{}

\bibitem[Babic et~al., 2021]{interpretibleAI-2021}
Babic, B., Gerke, S., Evgeniou, T., and Cohen, I.~G. (2021).
\newblock Beware explanations from {AI} in health care.
\newblock {\em Science}, 373(6552):284--286.

\bibitem[Benaglia et~al., 2009]{mixtools-2009}
Benaglia, T., Chauveau, D., Hunter, D.~R., and Young, D. (2009).
\newblock mixtools: An {R} package for analyzing finite mixture models.
\newblock {\em Journal of Statistical Software}, 32(6):1--29.

\bibitem[Benjamini and Hochberg, 1995]{benjaminihochberg-1995}
Benjamini, Y. and Hochberg, Y. (1995).
\newblock Controlling the false discovery rate: {A} new and powerful approach
  to multiple testing.
\newblock {\em J. Royal Statist.\ Soc.\ Series B}, 57:1289--1300.

\bibitem[Bollen, 1989]{bollen-lv-1989}
Bollen, K.~A. (1989).
\newblock {\em Structural Equations with Latent Variables}.
\newblock Wiley, New York.

\bibitem[Bretz et~al., 2016]{westfall-mc-2016}
Bretz, F., Hothorn, T., and Westfall, P. (2016).
\newblock {\em Multiple Comparisons using R}.
\newblock Chapman and Hall/CRC, London.

\bibitem[Dempster et~al., 1977]{dlr-em77}
Dempster, A., Laird, N., and Rubin, D.~B. (1977).
\newblock Maximum likelihood from incomplete data via the {EM} algorithm.
\newblock {\em J. Royal Statist.\ Soc., Series B}, 39(1):1--38.

\bibitem[Fawcett, 2006]{fawcett-roc-2016}
Fawcett, T. (2006).
\newblock An introduction to {ROC} analysis.
\newblock {\em Pattern Recognition Letters}, 27(8):861--874.

\bibitem[Hastie et~al., 2001]{friedmanhastietibshirani-2001}
Hastie, T., Tibshirani, R., and Friedman, J. (2001).
\newblock {\em The Elements of Statistical Learning: Data Mining, Inference,
  and Prediction}.
\newblock Springer--Verlag, New York.

\bibitem[Holtgrewe, 2010]{fu_mi_publications962}
Holtgrewe, M. (2010).
\newblock Mason: A read simulator for second generation sequencing data.
\newblock {\em Technical Report FU Berlin}.

\bibitem[Karr et~al., 2021a]{markovstructure-2021}
Karr, A.~F., Hauzel, J., Porter, A.~A., and Schaefer, M. (2021a).
\newblock {Markov Structure of Genomes, with Application to Outlier
  Identification, Read Classification and Detection of Contamination}.
\newblock Technical report, Fraunhofer Center Mid-Atlantic, Riverdale, MD.

\bibitem[Karr et~al., 2021b]{dqdegradation-2021}
Karr, A.~F., Hauzel, J., Porter, A.~A., and Schaefer, M. (2021b).
\newblock {Measuring Quality of DNA Sequence Data via Degradation}.
\newblock Technical report, Fraunhofer Center Mid-Atlantic, Riverdale, MD.

\bibitem[Karr et~al., 2019]{recordlinkage-plos1-2019}
Karr, A.~F., Taylor, M.~T., West, S.~L., Setoguchi, S., Kou, T.~D., Gerhard,
  T., and Horton, D.~B. (2019).
\newblock A comparison of record linkage software and algorithms using
  real-world data.
\newblock {\em PLoS ONE}, 14(9):e0221459.

\bibitem[McLachlan and Peel, 2000]{mclachlan-mixture-2000}
McLachlan, G.~J. and Peel, D. (2000).
\newblock {\em Finite Mixture Models}.
\newblock Wiley, New York.

\bibitem[Pearl and Mackenzie, 2018]{bookofwhy-2018}
Pearl, J. and Mackenzie, D. (2018).
\newblock {\em The Book of Why: The New Science of Cause and Effect}.
\newblock Basic Books, New York.

\bibitem[{R Core Team}, 2020]{R-2021}
{R Core Team} (2020).
\newblock {\em R: A Language and Environment for Statistical Computing}.
\newblock R Foundation for Statistical Computing, Vienna, Austria.

\bibitem[Sacks et~al., 1989]{sacks-etal-computerexperiments-1989}
Sacks, J., Welch, W.~J., Mitchell, T.~J., and Wynn, H.~P. (1989).
\newblock Design and analysis of computer experiments.
\newblock {\em Statistical Science}, 4(4):309--435.

\bibitem[Santner, 2003]{santner-computerexperiments-2003}
Santner, T. (2003).
\newblock {\em The Design and Analysis of Computer Experiments}.
\newblock Springer--Verlag, Berlin.

\bibitem[Westfall and Young, 1993]{westfallyoung93}
Westfall, P.~H. and Young, S.~S. (1993).
\newblock {\em Resampling-Based Multiple Testing: {E}xamples and Methods for
  {$p$}-Value Adjustment}.
\newblock Wiley, New York.

\end{thebibliography}

\end{document}